\documentclass[11pt,a4paper]{article}
\usepackage{jheppub}
\usepackage{amsmath,amssymb,amsfonts}
\usepackage{mathrsfs}
\usepackage{bbm}
\usepackage{slashed}
\usepackage{graphicx}
\usepackage{verbatim}
\usepackage[T1]{fontenc}
\usepackage[utf8]{inputenc}
\notoc
\begin{document}

\title{Non-linear black hole dynamics and Carrollian fluids}

\author[a, b, c]{Jaime Redondo--Yuste,}
\author[b]{and Luis Lehner}
\affiliation[a]{Niels Bohr International Academy, Niels Bohr Institute, Blegdamsvej 17, 2100 Copenhagen, Denmark}
\affiliation[b]{Perimeter Institute for Theoretical Physics, 31 Caroline Street North, Waterloo, ON N2L 2Y5, Canada}
\affiliation[c]{Department of Physics and Astronomy, University of Waterloo, Waterloo, Ontario, N2L 3G1,
Canada}
\emailAdd{jaime.redondo.yuste@nbi.ku.dk}
\emailAdd{llehner@perimeterinstitute.ca}
\arxivnumber{2212.06175}

\abstract{
The dynamics of black hole horizons has recently been linked to that of Carrollian
fluids. This results in a dictionary between geometrical quantities and those
of a fluid with unusual properties due its underlying Carrollian symmetries.
In this work we explore this relation in dynamical settings with the interest of 
shedding light on either side by relevant observations. In particular: we discuss
how the null surface where the Carrollian fluid evolves is affected by its behavior;
that the fluid's equilibration properties are tied to teleological considerations; the connection of higher derivative contributions as both source of energy and dissipation for the fluid and the non-linear behavior of black holes. This latter point, connects with discussions of non-linear modes
in the relaxation to equilibrium of perturbed black holes.
}

\maketitle

\section{Introduction}
The understanding of gravity around black holes has long been the subject of scrutiny. These objects have been a tremendous source of inspiration and central to questions ranging from astrophysics to field theories. At the same time, they have long presented puzzles that have often benefited from insights from other branches of physics. In particular, the membrane paradigm~\cite{damour_black-hole_1978, Thorne:1986iy, price_membrane_1986} connecting the physics of black holes with that of relativistic fluids at a timelike surface in the vicinity of the event horizon
shed light on particular aspects of the region's behavior.
However, freedom in the choice of such surface could obscure somewhat particular connections. In the context of black holes in asymptotically AdS spacetimes, this issue was largely addressed by choosing the AdS boundary --a unique surface in the spacetime-- and a clean fluid/gravity correspondence was
established~\cite{kovtun_holography_2003,Baier:2007ix,Bhattacharyya:2007vjd,VanRaamsdonk:2008fp}. This enabled, in particular, motivating the possibility of turbulent behavior in General Relativity with a negative cosmological constant~\cite{VanRaamsdonk:2008fp,carrasco_turbulent_2012,Adams:2013vsa}. In the opposite direction, insights on possible completion of relativistic hydrodynamics and the analysis of transport coefficients, have been discussed from the gravitational side (e.g.~\cite{Heller:2007qt,Cremonini:2011iq,deBoer:2015ija})

In the context of asymptotically flat spacetimes special surfaces to attempt a similar connection are future null infinity $\mathscr{I}^+$ or the black hole event horizon. Since both are null surfaces, special care must be taken to deal with their singular nature. In the case of the event horizon, such a program has recently been carried out, leading to the discovery that gravity around the black hole is dual to Carrollian fluids~\cite{donnay_carrollian_2019}. These fluids appear as the limit of relativistic hydrodynamics as the speed of light goes to zero. Even though in that limit a single particle is not allowed to have any dynamics~\cite{Duval:2014uva}, interacting particles retain some dynamics in the limit~\cite{Bergshoeff:2014jla, Bergshoeff:2022eog}. The physical nature of this Carrollian hydrodynamics remains puzzling; in particular
the notion of entropy and a microscopic description of such fluids are still work in progress (e.g.~\cite{deBoer:2017ing,Petkou:2022bmz}). 
In the current work, we further explore this connection by focusing on the dynamical aspects. In particular, we study the black hole region's behavior both geometrically and in terms of a (Carrollian) fluid description. To this end, we focus on the relaxation to equilibrium of a black hole that is
perturbed by a pulse of gravitational waves --physically motivated by scenarios describing the after-merger stage of binary black hole collisions. {We find that the non--linear dynamics results in an excitation of shorter wavelength modes which occurs
prior to the perturbation --related to the teleological character of the horizon. Depending on the strength of the perturbation, appreciable non-linear energy exchange is
induced and short wavelength modes can become even more relevant than those modes excited by the perturbation. To understand this behavior, we analyze the structure of the equations and note the Carrollian fluid dynamics has a non-trivial/non-linear relation with the ``teleological'' volume (the horizon's surface) it occupies. To unearth this effect, we connect to the bulk spacetime thus revealing this intricate connection. In particular, this allows for motivating a definition of gravitational Reynolds number to help distinguish 
between laminar and turbulent regimes.} Further, since our studies on this topic are intrinsically linked with the behavior of horizons, 
and in particular their non-linear behavior, we draw from also a 
complementary way to explore dynamical horizons. Particularly, the 
transition from non-linear to linear regimes.

This work is organized as follows: section~\ref{sec:geometry} revisits the geometry of null surfaces and how can they naturally be endowed with a Carrollian structure. Section~\ref{sec:duality} establishes the gravitational dynamics of this geometry, and its mapping to   Carrollian fluids by properly identifying geometric and fluid degrees of freedom. In section~\ref{sec:perturbations} we describe our non--linear perturbative set--up and the equations that will determine the horizon's dynamics. We introduce the quantities that we will monitor during the evolution in section~\ref{sec:monitor} and discuss our numerical implementation in section~\ref{sec:numerics}. Finally we analyze the results of our simulations in section~\ref{sec:results}. We end with final observations in section~\ref{sec:conclusions}.  We also include in the appendices more
details on Carrollian geometry as well as
further potential connections with hydrodynamical behavior that would be explored elsewhere.

In what follows, we restrict to 3+1 spacetime dimensions, indices with upper case letters are intrinsic, 2D, indices $A, B = 2, 3$ and lower case ones denote spacetime indices $a, b = 0 \dots 3$.

\section{Geometry of null surfaces}\label{sec:geometry}

Let ${\bf H}$ be a null surface, and assume it is located at some adapted coordinate $\rho = 0$. It is always possible to find a chart containing the surface such that the metric is written in the form~\cite{Gomez:2000kn,chrusciel2020geometry}:
\begin{align}\label{eq:metric-gauge}
    ds^2 = - V dv^2 + 2dvd\rho + 2\Upsilon_A dv dx^A + \gamma_{AB}dx^A dx^B,
\end{align}
where $v$ is the advanced null time, $\rho$ parameterizes the ``distance'' to the surface, and $x^A$ are the angular coordinates on the sphere, $A, B=2,3$. We expand the metric in powers of $\rho$,
\begin{equation}\label{eq:metric-expansion}
    \begin{split}
        V&=2\kappa\rho + O(\rho^2), \quad \Upsilon_A = U_A \rho + O(\rho^2), \quad  \gamma_{AB} = \Omega_{AB}-2\lambda_{AB}\rho + O(\rho^2),
    \end{split}
\end{equation}
where $\kappa$, $U_A$, $\Omega_{AB}$ and $\lambda_{AB}$ are functions of 
$(v,x^A)$. Note this gauge is a restricted version of the one chosen in~\cite{adami_null_2021}; their geometrical characterization also applies for our 
choice, upon making the necessary identifications. We next define the geometric variables that will be used to characterize the evolution of the null surface.

The null surface has a null normal vector $l^a$ that is also tangent to it. We introduce another null normal vector to the surface, $n^a$, normalized such that $n^a l_a = -1$. These
two vectors are given by 
\begin{align}\label{eq:null_vectors}
    l_a dx^a = -\frac{1}{2}Vdv+d\rho, \quad n_adx^a = -dv.
\end{align}
It is useful to characterize the geometry of the null surface in terms of quantities whose definition do not depend on the coordinates, but that acquire a simple expression in terms of our coordinate choice. Notice that the induced metric on the surface ${\bf H}$ is given by $h_{ab}=g_{ab}+2l_{(a}n_{b)}$, which yields
\begin{align}\label{eq:induced-metric}
    ds^2_{\bf H} = \Omega_{AB}dx^A dx^B.
\end{align}
Therefore, for each null normal $m = \{l_a, \, \, n_a\}$, which we debote with a upperscript in between brackets, we can construct its deviation tensor. This is defined as the projection onto the null surface of its covariant derivative 
\begin{align}\label{eq:deviation-tensor}
    B^{(m)}_{AB} = \left(h^a_A h^b_B \nabla_b m_a\right)\lvert_{\rho = 0} = N^{(m)}_{AB} + \frac{1}{2}\theta^{(m)} \Omega_{AB},
\end{align}
that we decompose into its trace (the expansion) and its traceless symmetric part (the shear).\footnote{The deviation tensor is completely symmetric since we are evaluating it at a surface. Any surface can be described by a locally hypersurface orthogonal congruence and therefore, its twist (the antisymmetric part of the deviation tensor) vanishes~\cite{Wald:1984rg}.} These are given now by 
\begin{equation}\label{eq:expansion-shear-def}
    \begin{split}
        \theta^{(l)} &= \frac{\Dot{\Omega}}{2\Omega}, \quad \theta^{(n)} = \Omega^{AB}\lambda_{AB}, \\
        N^{(l)}_{AB}&= \frac{1}{2}\Dot{\Omega}_{AB}-\frac{\theta^{(l)}}{2}\Omega_{AB}, \quad N^{(n)}_{AB}= \lambda_{AB}-\frac{\theta^{(n)}}{2}\Omega_{AB},
    \end{split}
\end{equation}
where the dot ($\dot{}$) denotes the coordinate derivative with respect to $v$, and $\Omega = \textrm{det}\, \Omega_{AB}$. Finally, the last variable that we will use is the Hajiceck one-form, which is given by 
\begin{equation}\label{eq:hajiceck-def}
    \mathcal{H}_A = \left(h^a_A l_b \nabla_a n^b\right)\lvert_{\rho = 0} = -\frac{1}{2}U_A.
\end{equation}
We can provide a geometric meaning to all the quantities involved in the metric expansion~\eqref{eq:metric-expansion}. The parameter $\kappa$ is the non-affinity of our parameterization, which coincides with the surface gravity for stationary black holes since $l^a\nabla_a l^b = \kappa l^b$. The shift term $U_A$ is related to the Hajiceck one-form. The angular metric $\Omega_{AB}$ describes the intrinsic metric of the null surface, whereas the term $\lambda_{AB}$ describes the angular metric right outside of the surface itself, and therefore takes into account ambient effects. 

The geometry of a null surface can be endowed with a Carrollian structure. This observation comes from the realization that the diffeomorphisms $\xi$ that preserve the gauge~\eqref{eq:metric-gauge}, once restricted to the null surface, are just 
\begin{equation}
    \xi = f(v, x^A)\partial_v + Y^A(x^A) \partial_A,
\end{equation}
which are the infinitesimal version of the Carrollian diffeomorphisms~\cite{donnay_extended_2016}. 

A Carroll structure is given by a surface ${\bf H}$, a degenerate metric on that surface $h_{AB}$ and a preferred generator, $\ell^A = h^{A}_bl^b$ such that $l^A h_{AB}=0$. Although this structure can be defined intrinsically for any given null surface (e.g.~\cite{Hartong:2015xda,Ciambelli:2019lap,Freidel:2022bai}, it is perhaps easier to make the identification by comparing the near-horizon geometry~\eqref{eq:metric-gauge} with the Randers--Papapetrou parameterization~\cite{ciambelli_paving_nodate}, given by 
\begin{equation}\label{eq:Randers-Papapetrou-limit}
    \begin{split}
        ds^2 &= -c^2 \alpha^2 dt^2 + 2 c^2 \alpha b_A dt dx^A + 
        \\ &+ (\Omega_{AB}-c^2 b_A b_B) dx^A dx^B\xrightarrow{c^2 \, \to \, 0} \Omega_{AB}dx^A dx^B.
    \end{split}
\end{equation}
Then, the null surface of interest ${\bf H}$ is endowed with a Carroll structure with degenerate geometry $\Omega_{AB}$ and the following identifications:
\begin{equation}\label{eq:Carrollian-geom-identification}
    \alpha = \sqrt{2\kappa}, \quad b_A = \frac{1}{\sqrt{2\kappa}} U_A,
\end{equation}
with $\alpha$ the lapse function and $b_A$ the shift or temporal connection. More importantly, the distance to the surface $\rho$ plays the role of the speed of light $c^2$. This motivates the following intuition: A timelike surface located at some distance $\rho$ to the horizon has a relativistic fluid dual as per the usual membrane paradigm~\cite{damour_black-hole_1978, price_membrane_1986, Thorne:1986iy}. Therefore, as one takes the limit in which the surface becomes null $\rho \to 0$, then one must take the ultrarrelativistic or Carrollian limit $c^2 \to 0$ on the dual fluid. 

\section{Carrollian membrane paradigm}\label{sec:duality}

\subsection{Dynamics of null surfaces}

The membrane paradigm~\cite{damour_black-hole_1978} identifies the gravitational dynamics on the stretched horizon to the evolution of a relativistic fluid in $2+1$ dimensions. The stretched horizon is defined to be a timelike surface, sitting at some distance outside of the event horizon of a given black hole spacetime. If one insists on finding a duality with the dynamics of a relativistic fluid at the level of the event horizon, the quantities involved in this process diverge. These can be regularized by introducing a family of fiducial inertial observers~\cite{Thorne:1986iy}. However, this procedure depends explicitly on the foliation of stretched horizons chosen, and therefore it is not universal. One can avoid the divergences by identifying the degrees of freedom at the level of the event horizon with those of a Carrollian fluid, as it was shown in~\cite{donnay_carrollian_2019}.  In this section we frame this construction in a different way, not requiring  the use of a foliation of stretched horizons.

The well-defined gravitational action of a spacetime $\mathcal{M}$ with a null boundary $\bf{H}$ is given by \cite{lehner_gravitational_2016}:
\begin{equation}
    S = \int_{\mathcal{M}} d^4 x \sqrt{-g} \, R + \int_{{\bf H}}d^2x \, dv \, \sqrt{-\Omega}\, \kappa
\end{equation}
where we have fixed units such that $8\pi G = 1$. The variations with respect to the boundary degrees of freedom were shown in~\cite{Chandrasekaran:2021hxc} to be equivalent to the conservation of the null stress-energy tensor given by 
\begin{equation}\label{eq:def-set-null}
    T^a_b = - \left(W^a_b - W\delta^a_b\right),
\end{equation}
where $W^a_b = h^a_\alpha h^\beta_b \nabla_\beta l^\alpha$, where $l^a$ is the null generator of the surface, and $W = W^a_a$ is its trace. Notice how this stress energy tensor is similar in structure and in construction to the Brown--York stress energy tensor~\cite{York:1986it,Brown:1992br}, although that is constructed for a timelike surface.  A discussion about the relation between this stress energy tensor and the null limit of a timelike surface is given in \cite{Chandrasekaran:2021hxc}. Also, from the full spacetime point of view, this stress energy
tensor is obtained through a suitable projection with $h_{ab}$. This projector is not unique but is invariant with respect of scalings, see~\cite{Freidel:2022vjq}. Regardless,
once a pair of null vectors $\{l,n\}$is defined a consistent decomposition is made, 
and employed to study a black hole's behavior and explore its dynamics from several angles as we do in this work.

Gravitational dynamics on the null boundary are given by the conservation of this null stress energy tensor:
\begin{align}\label{eq:conservation-set}
    \nabla_b T^b_a = - G h_{ab} R^{bc} l_c = 0,
\end{align}
where $R_{ab}$ is the Ricci tensor and we are assuming a vacuum spacetime. These equations give directly the Raychaudhuri equation (the $b = 0$ component) and the Damour--Navier--Stokes equation (the $b=2,3$ components), also referred to sometimes as the Hajiceck equation~\cite{price_membrane_1986}.~\footnote{The $b=1$ component is trivial once projected onto the horizon.} These equations are given, in terms of the geometric variables, by 
\begin{equation}\label{eq:Ray-Damour-eqs}
    \begin{split}
        \Dot{\theta^{(l)}} - \kappa\theta^{(l)} + \frac{1}{2}(\theta^{(l)})^2 + N^{(l)}_{AB}N^{AB \, (l)}&=0, \\
        \Dot{\mathcal{H}}_A + \theta^{(l)} \mathcal{H}_A - \nabla_A \kappa - \frac{1}{2}\nabla_A\theta^{(l)} + \nabla^B N^{(l)}_{AB}&=0,
    \end{split}
\end{equation}
where~$\nabla_A$ here is the covariant derivative associated to the metric~$\Omega_{AB}$. Notice that these equations are just a restricted version of the constraint equations derived in~\cite{adami_null_2021}, which also allow for a geometric description~\cite{Gourgoulhon:2005ng}. By defining everything from the variational principle adapted for a null surface we have avoided introducing a foliation of stretched horizons. 

\subsection{Mapping to the Carrollian fluid}

A Carrollian fluid can be described as the $c\to 0$ limit of a relativistic fluid~\cite{ciambelli_covariant_2018,Freidel:2022bai}. Upon taking the limit, the fluid is described by a similar set of hydrodynamical variables: the energy density $e$, the pressure $p$, a pair of heat currents $\mathcal{J}^A$ and $\pi^A$, and a pair of dissipation tensors $\Sigma^{AB}$ and $\Pi^{AB}$. It could seem that by introducing another pair of heat current and dissipation tensors we are increasing the number of degrees of freedom of the system. This is not the case, since these two additional terms enter in a different order in the expansion in powers of $c$ of the fluid, and must satisfy two constraints:
\begin{equation}\label{eq:Carrollian-constraints}
    \begin{aligned}
        \Sigma_{AB}N^{(l) \, AB} &= 0, \\
        \Dot{\mathcal{J}}^A + \theta^{(l)}\mathcal{J}^A + (\hat{\nabla}_B + \varphi_B)\Sigma^{AB} &= 0.
    \end{aligned}
\end{equation}
Here $\theta^{(l)}$ and $N^{(l)}_{AB}$ are the expansion and shear of the Carroll structure where the fluid is defined.~\footnote{These coincide with the expansion and shear of the ``velocity'' of the Carrollian fluid, if written from the point of view of a stress energy tensor, which is given by the null generator $\ell^A$ of the horizon.} The dot denotes the time derivative $\Dot{y} = \partial_v y$. We have also introduced the Carrollian acceleration~\cite{ciambelli_covariant_2018}
\begin{equation}
    \varphi_A = \frac{1}{2\kappa}\left(\frac{\Dot{\kappa}}{\kappa}\mathcal{H}_A - 2 \Dot{\mathcal{H}}_A + \partial_A\kappa\right),
\end{equation}
and the Carrollian derivative $\hat{\nabla}_A$ (see appendix~\ref{App:Carrollian}). Then, the Carrollian fluid is described by the following two evolution equations
\begin{equation}\label{eq:Carrollian-evolution}
    \begin{aligned}
        &\Dot{e} + \theta^{(l)}e +\alpha(\hat{\nabla}_A + 2\varphi_A)\mathcal{J}^A + \left(N_{AB}^{(l)}+\frac{1}{2}\theta \Omega_{AB}\right)(\Pi^{AB} + p \Omega^{AB}) = 0, \\
        &\Dot{\pi}_B + \theta^{(l)}\pi_B + \alpha\left(\hat{\nabla}_A+\varphi_A\right)(\Pi^A_B + p \delta^A_B) +\alpha\varphi_B e - \varpi^{AB}\mathcal{J}_A = 0.
    \end{aligned}
\end{equation}
In the above $\varpi_{AB}$ is the Carrollian vorticity (see next section and appendix~\ref{App:Carrollian}). Now one realizes that these equations are just the gravitational equations~\eqref{eq:Ray-Damour-eqs} in disguise. Identifying terms is straightforward, and yields the following fluid/gravity dictionary:
\begin{equation}\label{eq:dictionary}
    \begin{aligned}
        e &= \theta^{(l)}, \quad p = -\kappa , \\
        \Pi_{AB} &= 2\eta N_{AB}^{(l)} + \zeta \theta^{(l)}\Omega_{AB}, \quad \eta = \frac{1}{2}, \quad \zeta = -\frac{1}{2}, \\
        \pi_{A} &= -\frac{1}{\kappa}\left(\nabla_A\kappa  - \mathcal{H}^B \Dot{\Omega}_{AB} - \frac{1}{\kappa}\mathcal{H}_A \Dot{\kappa}\right), \quad
        \Sigma^{AB} = \mathcal{J}^A = 0.
    \end{aligned}
\end{equation}
In obtaining this dictionary, notice that the second pair of heat flux and dissipation tensor $\Sigma_{AB} = \mathcal{J}_A = 0$ vanish. The vanishing of $\mathcal{J}_A$ is due to the identification with the Raychaudhuri equation, and then the constraints~\eqref{eq:Carrollian-constraints} necessarily imply the vanishing of $\Sigma_{AB}$. A very similar result is obtained by mapping the gravitational stress energy tensor~\eqref{eq:def-set-null} to the stress energy tensor of a Carrollian fluid~\cite{Freidel:2022bai, Freidel:2022vjq}. The main difference lies on the ambiguity in assigning contributions to pressure and the bulk viscosity $\zeta$ term of the dissipation. Their dissipation tensor is required to be traceless, which implicitly assumes vanishing bulk viscosity. This is the case, e.g. in the fluid/gravity correspondence for asymptotically AdS spacetimes, where the fluid retains a conformal symmetry. However in the case of the horizon there are no symmetry arguments which would result in a vanishing bulk viscosity. Therefore, one would expect, in a similar fashion as in the identifications done in the membrane paradigm works~\cite{Thorne:1986iy}, a non--vanishing bulk viscosity, thus we adopt this point of view. 

Our result also yields a very similar dictionary to the one first proposed in~\cite{donnay_carrollian_2019}, but with some terms that depend on the gradients of the surface gravity absent. The difference was discussed in~\cite{Chandrasekaran:2021hxc}, where it was explained how those terms appear as additional terms due to the regularization needed to take the null limit, when starting from a timelike surface. Since we do not consider a timelike stretched horizon at any time, these terms do not appear. 

The resulting correspondence maps the dynamics on a black hole horizon to those of a Carrollian fluid with {\em very particular properties}. Importantly, we highlight the pressure is negative, and the energy density vanishes in equilibrium. At this point we find it relevant to note that there is an ambiguity in the identification: we could have chosen to re-define the dictionary~\eqref{eq:dictionary} with a negative energy density, positive pressure and positive bulk viscosity. Notice that the sign of the shear viscosity $\eta$ would remain unchanged. However, even though a fluid with negative pressure poses a challenging question about its interpretation it has a precedent in the membrane paradigm~\cite{damour_black-hole_1978}. On the other hand a negative energy density seems rather unphysical. A thermodynamical formalism for Carrollian fluids could help clarify this, as well as the question on the entropy current for Carrollian fluids. \\

Before ending this section, we highlight an important subtlety. While it is clear that a horizon can be mapped to a Carrollian fluid, in the dynamical regime, there are ingredients of such mapping that are yet to be clarified and which are inescapably linked to the very volume where such a fluid ``lives''. That is, its backreaction on it, as well as the influence of the embedding spacetime are important to close the system and studying its behavior. Furthermore, for the specific case of a horizon, the Carrollian fluid's dynamics
is also strongly determined by the teleological character of this
surface.  In particular, the requirement that in the asymptotic future $\theta^{(l)} \rightarrow 0$ which, in turn, implies $\theta^{(l)} > 0$ at early times in the dynamical case.
In what follows, we perform a perturbative analysis of the
spacetime in the neighborhood of the horizon to illustrate this relation and motivate further connections.

\section{Perturbative set--up}\label{sec:perturbations}

So far, the previous analysis has been purely kinematical and the established
dictionary, while interesting in its own right can potentially be misinterpreted
if translating from fluid considerations lightly. Namely, in
the case of fluids {\em dissipative effects} are encoded by the shear and expansion
appearing in the tensor $\Pi_{AB}$; however, for Carrollian fluids dual to
black hole horizon these need not be the case as the horizon is affected
by energy accretion.  Such process certainly affects the resulting fluid
dynamics in a non-linear fashion, which is not straightforwardly appreciated
in the equations~\eqref{eq:Carrollian-evolution}. To elucidate this behavior, we here consider
a horizon perturbed by incoming gravitational waves, and treat the system
perturbatively to sufficiently high order to keep track of its behavior. 
We note that in full non-linearity one should study this problem by 
tackling the full set of equations in a framework making contact with the horizon's behavior. A transparent way to this end  was presented 
in~\cite{Gomez:2000kn,Gomez:2002ev}. Alternatively, insights can be gained 
through the dynamical horizon approach~\cite{ashtekar_dynamical_2013,Ashtekar:2021kqj} or via
a suitable second order perturbative approach~\cite{Loutrel:2020wbw,Ripley:2020xby}. However,
we here want to keep a close connection with the Carrollian fluid structure and
find it convenient to proceed as follows.

We will look at a simplified situation: a spherically symmetric black hole perturbed by a gravitational wave.\footnote{For the rotating case, a starting point would make use of the metric in, e.g.~\cite{Gray:2022svz}.} The geometry will be described by $g_{ab} = g_{ab}^{\textrm{Sch}} + \epsilon h_{ab} + \epsilon^2 h^{(2)}_{ab}$, where $\epsilon \ll 1$ is some perturbative parameter and $g_{ab}^{\textrm{Sch}}$ the Schwarzschild metric, given by 
\begin{equation}
    g_{ab}^{\textrm{Sch}} = -\frac{\rho}{4m}dv^2+2dvd\rho+(4m^2 +  4m\rho)q_{AB}dx^A dx^B,
\end{equation}
where $q_{AB}$ is the usual unit sphere metric. For consistency, the perturbative
expansion is carried to ${\cal O}(2)$ as the `energy' $e \equiv \theta^{(l)}$ is
a quantity of that order outside equilibrium. The perturbation is also expanded in the gauge~\eqref{eq:metric-gauge}, so we define 
\begin{equation}
    \begin{aligned}
        \kappa &= \frac{1}{4m}(1 + \epsilon k + \epsilon^2 K), \quad 
        U_A = -2(\epsilon h_A + \epsilon^2 H_A), \\
        \Omega_{AB} &= 4m^2(q_{AB} + \epsilon c_{AB} + \epsilon^2 C_{AB}), \quad \lambda_{AB} = 2m(-q_{AB} + \epsilon s_{AB} + \epsilon^2 S_{AB}),
    \end{aligned}
\end{equation}
where all terms $(k,h_A,c_{AB},s_{AB}; K, H_A, C_{AB}, S_{AB})$ are functions of $(v,x^A)$ and we assume that the angular part is traceless to each order, $q^{AB}c_{AB}= q^{AB} C_{AB} = q^{AB}s_{AB} =  q^{AB}S_{AB} = 0$. The source of perturbations is encoded by
the trace-free (first and second order in $\epsilon$ tensors $\{s_{AB},S_{AB}\}$.
They encode perturbations incoming from the bulk. We will define these perturbations 
with a suitable, physically motivated, profile. The rest of the perturbation degrees of freedom are fixed by solving the Einstein equations in vacuum $R_{ab} = 0$. We observe that the $R_{vv}$ and $R_{vA}$ components give just the Raychaudhuri and Damour equations~\eqref{eq:Ray-Damour-eqs}. Moreover, as it was observed in~\cite{Gomez:2000kn,adami_null_2021}, the $R_{\rho b}$ components constrain the higher order terms in the $\rho$ expansion of the metric. Finally, we split the $R_{AB}$ equations into its trace and trace-free symmetric parts:
\begin{equation}\label{eq:full-GR-eqs}
    \begin{aligned}
    \Dot{\theta}^{(n)} + \kappa\theta^{(n)} + \theta^{(l)} \theta^{(n)} - (\nabla_A\mathcal{H}^A + \mathcal{H}_A\mathcal{H}^A)+\frac{R}{2}&=0, \\
    2\Dot{N}^{(n)}_{AB} - 4N^{C \,(n)}_{(A} N^{(l)}_{B)C}+\theta^{(n)} N^{(l)}_{AB} + (2\kappa-\theta^{(l)})N^{(n)}_{AB} &+ \\
    +R_{AB}^{\textrm{T-F}}-2(\mathcal{H}_A\mathcal{H}_B)^{\textrm{T-F}} - 2(\nabla_{(A}\mathcal{H}_{B)})^{\textrm{T-F}}&=0,
    \end{aligned}
\end{equation}
where $\nabla_A$ is the covariant derivative associated to the metric $\Omega_{AB}$, and $R_{AB}$ and $R$ are the Ricci and scalar curvatures associated to that connection. The above geometric variables are the ones defined in section~\ref{sec:geometry}, associated to the choice of null basis~\eqref{eq:null_vectors} The T-F superscript denotes the trace-free symmetric part. 
We now expand all relevant terms up to second order in $\epsilon$.  First, for convenience, we write the Ricci scalar as $R = 1 / (2m^2) +\epsilon R_1 + \epsilon^2 R_2$. The expansions and shears can then be expressed as,
\begin{equation}
    \begin{aligned}
    \theta^{(l)} &= \frac{\dot{\Omega}}{2\Omega} = \frac{1}{2}\Omega^{AB}\dot{\Omega}_{AB} = -\frac{\epsilon^2}{2}c^{AB}\dot{c}_{AB}, \\
    \theta^{(n)} &= \Omega^{AB}\lambda_{AB} = -\frac{1}{2m}(2 + \epsilon^2 X), \\
    N^{(l)}_{AB} &= \frac{\dot{\Omega}_{AB}}{2}-\frac{\theta_l}{2}\Omega_{AB} = 2m^2\epsilon\left[\dot{c}_{AB} + \epsilon \left(\dot{C}_{AB} + \frac{1}{2} c^{CD}\dot{c}_{CD}q_{AB}\right)\right], \\
    N^{(n)}_{AB} &= \lambda_{AB} - \frac{1}{2}\theta_n \Omega_{AB} = 2m\epsilon\left[c_{AB} + s_{AB} + \frac{1}{2}\epsilon \left(X q_{AB} + 2C_{AB}\right)\right].
    \end{aligned}
\end{equation}
with $X = c^{AB}(c_{AB} + s_{AB})$. We can then re-express Einstein equations and
solve for each relevant order, obtaining:
From the trace of $R_{AB}$:
\begin{equation}
    \begin{aligned}
    k &= -\nabla^A h_A + 2m^2R_1, \\
    K &= -(\nabla^A H_A + h^Ah_A) + 2m c^{AB}\dot{c}_{AB} - \frac{1}{2}X - 2m\dot{X} + 2m^2R_2.
    \end{aligned}
\end{equation}
and from the trace-free part of $R_{AB}$:
\begin{equation}\label{eq:evol-c-C}
    \begin{aligned}
    \dot{c}_{AB} &= -2\dot{s}_{AB} - \frac{1}{2m}(c_{AB} + s_{AB}) + \frac{1}{2m}(\nabla_Ah_B + \nabla_B h_A - \nabla^C h_C q_{AB}), \\
    \dot{C}_{AB} &= -\dot{X}q_{AB} + 2(c^C_{(A} + s^C_{(A})\dot{c}_{B)C} -\frac{1}{4m}[Xq_{AB} + 2C_{AB} + 2k(c_{AB}+s_{AB})] +\frac{1}{2}c^{CD}\dot{c}_{CD}q_{AB}\\
    &\quad  + \frac{1}{m}\left(h_Ah_B-\frac{1}{2}h^Ch_C q_{AB}\right) + \frac{1}{2m}\left(2\nabla_{(A}H_{B)}- \nabla^C H_C q_{AB} + c^{CD}\nabla_C h_D q_{AB} \right.  - \left.\nabla^C h_C c_{AB} \right),
    \end{aligned}
\end{equation}
These equations make the connection of quantities intrinsic to the horizon and the behavior of the embedding space evident. In particular, this implies the Carrollian fluid dynamics can be conveniently interpreted from this point of view.
For instance, to leading order,
\begin{eqnarray}\label{eq:viscous}
\Pi_{AB} &=& 2 m^2 \left[ - 2\dot s_{AB} -\frac{(c_{AB} + s_{AB})}{2m}  + \frac{\sigma^H_{AB}}{2m}  \right] \nonumber \\
&=&  \left[ - 4 m^2 \dot s_{AB} - \frac{N^{(n)}_{AB}}{2}  + 2m \, \sigma^H_{AB}  \right]
\end{eqnarray}
with $2\,\sigma^H_{AB}=(\nabla_Ah_B + \nabla_B h_A - \nabla^C h_C q_{AB})$,  the shear associated to the Hajicek field and in the second line we replaced to
leading order with the shear associated to $n^a$. Clearly, $\Pi_{AB}$ contains contributions of dissipation as well as source of energy. As well, the
equation of state to leading order results:
\begin{equation}
p = - \frac{1}{4m} \left( 1 - D_A h^A + 2 m^2 D_A D_B c^{AB} \right ) \, ,    
\end{equation}
where $D_A$ is the covariant derivative associated to the metric on the sphere $q_{AB}$. Notice that for any tensorial quantity $T$ we can write $\nabla_A T = D_A T + \epsilon \, \Gamma \, T+\dots$ for some terms $\Gamma$ depending on derivatives of $c_{AB}$. The above equation indicates that the equation of state depends on both the  metric of spheres at $v=const$ and divergence of the Hajicek field.

The evolution equations, to order two in $\epsilon$, is given by Eq.~\eqref{eq:evol-c-C} and the following:
\begin{equation}\label{geometriceqns}
    \begin{aligned}
    \dot{\vartheta} &= \frac{1}{4m}\vartheta - m^2\dot{c}^{AB}\dot{c}_{AB}, \\
    \dot{h}_A &= \frac{1}{4m}\nabla_A k -\frac{1}{2}\nabla^B \dot{c}_{AB}, \\
    \dot{H}_A &= \frac{1}{4m}\nabla_A K + \nabla_A \vartheta - \frac{1}{2}\nabla^B \dot{C}_{AB} + \frac{1}{2}c^{BC}\nabla_C \dot{c}_{AB}.
    \end{aligned}
\end{equation}
where we have introduced the variable $\vartheta = \theta^{(l)}$, since it is the only expansion that appears explicitly in our system as a dynamical variable. We choose to express them in terms of 
spin-weighted quantities for convenience when implementing
them numerically. In order to do this, we introduce a complex dyad $q^A$ on the sphere such that $q_{AB}=q_{(A}\Bar{q}_{B)}$, and normalized with $q_A \Bar{q}^A = 2$, $q_A q^A = \Bar{q}_A \bar{q}^A = 0$. Then, the Hajiceck vector field is described by the complex function $\mathcal{H } = q^A\mathcal{H}_A$, and symmetric rank--2 tensors $T_{AB}$, $\{c_{AB}, C_{AB}, s_{AB}, S_{AB} \}$, by the complex function $2T = q^A q^B T_{AB}$. The covariant derivatives on the sphere can also be projected into this dyad to construct spin raising and lowering operators, dubbed $\eth$ and $\bar{\eth}$, as discussed in~\cite{newman_approach_1962, gomez_eth_1997}. 

To sum up, we started with a set of first order $\{k, h_A, c_{AB}\}$ and second order $\{K, H_A, C_{AB}\}$ dynamical degrees of freedom, and some source terms that describe perturbations incoming from the bulk $\{s_{AB}, S_{AB}\}$. We have used a subset of Einstein equations ($R_{AB}=0$) to eliminate algebraically the corrections to the surface gravity $k$ and $K$ in terms of other dynamical variables as well as derive evolution
equations for $\{c_{AB}, C_{AB}\}$. To leading order in $\epsilon$ and in spin-weighted form, the resulting equations are: 
\begin{equation}\label{eq:full-eqs-first}
    \begin{aligned}
        \dot{h} &= -\frac{1}{8m}(\eth^2 \bar{h} + \eth\bar{\eth}h) + \frac{1}{32m}(\eth\eth^2\bar{c}+\eth\bar{\eth}^2c) - \frac{1}{4}\bar{\eth}\dot{c}, \\
        \dot{c} &= -2\dot{s} - \frac{1}{2m}(c+s) + \frac{1}{2m}\eth h.
    \end{aligned}
\end{equation}
The second order terms satisfy the evolution equations:
\begin{equation}\label{eq:full-eqs-second}
    \begin{aligned}
        &\dot{\vartheta} = \frac{1}{4m}\vartheta - 2m^2\dot{c}\bar{\dot{c}}, \\
        &\dot{H} = \eth \vartheta -\frac{1}{4}(\eth\bar{\eth}H + \bar{\eth}\eth H + \eth^2 H - \bar{\eth}C) - \frac{1}{2}(h\eth \bar{h} + \bar{h}\eth h + h \bar{\eth}h) - \frac{1}{8}\bar{\eth}\left[(\eth\bar{h} + \bar{\eth}h)s\right] \\
        &+ \frac{1}{8}(h\eth^2\bar{c} + \bar{h}\eth\bar{\eth}c + \eth h \eth \bar{c} + \eth \bar{h}\bar{\eth}c) - \frac{1}{16}(\bar{\eth}h\bar{\eth}c +2 \bar{\eth}\bar{h}\eth c) -   \frac{1}{32}\left[2h \bar{\eth}^2 c + 2\bar{\eth}\bar{h}\eth c -(\bar{\eth}^2 c + \eth^2\bar{c})(\bar{\eth}c + \bar{\eth}s)\right] \\
        &+ \frac{1}{8}(3\dot{c}\eth\bar{c}+\dot{\bar{c}}\eth c) - \frac{1}{4}(2c\eth \bar{c} + 2\bar{c}\eth c + c\eth \bar{s} + \bar{s}\eth c) -  \frac{1}{4}( \dot{c}\eth \bar{s} + \bar{s}\eth \dot{c} + \dot{\bar{c}}\eth s + s \eth \dot{\bar{c}} + c\eth\dot{\bar{s}} + \dot{\bar{s}}\eth c + \bar{c}\eth\dot{s} + \dot{s}\eth\bar{c}  ), \\
        &\dot{C}= -C + \eth H + \frac{1}{2}(\eth \bar
        h + \bar{\eth}h)s - \frac{1}{8}(\bar{\eth}^2 c + \eth^2\bar{c})(c+s) + \frac{1}{4}(h\bar{\eth}c+\bar{h}\eth c).
    \end{aligned}
\end{equation}
For convenience, we set $2m = 1$; further, as we will perturb the system
in the long-wavelength regime, we drop terms containing third derivatives 
assuming a quasispherical regime of sorts~\cite{Bishop:1996gt}.  

In order to complete the description of the evolution of the system, we need to provide boundary conditions and a prescription for the incoming
gravitational wave, which is encoded in the variable $s$. For the latter, we give a compact support
pulse of gravitational waves, with frequency and decay time associated
to quasi-normal modes in Schwarzschild~\cite{Kokkotas:1999bd}.
For the former, we note that unless asymptotic boundary conditions
in the future are given such that $\vartheta \rightarrow 0$ the null surface will either expand to null infinity or converge towards the singularity. Therefore, we demand that $\vartheta, c, \mbox{and~} C$ are zero at asymptotic time $v\to\infty$. We complement this with choosing a vanishing Hajiceck field at $v\to\infty$, since we assume that asymptotically the metric is given by the Schwarzschild black hole in ``non-rotating'' coordinates. This amounts to fixing $h=H=0$ in the boundary.

\subsection{Non-linear considerations}
Before moving on to the numerical implementation of the relevant equations,
we note that resumming the equations for $\{\dot h_a,\dot H_A\}$ one can
see the non-linear structure of the Hajicek equation is, schematically, of the form:
\begin{equation}
    \begin{aligned}
        -4m\dot{H}_A &= \nabla^2H_A + \nabla_B\nabla_AH^B  + m^{-1} \left( H_A\nabla^BH_B + H_B\nabla^B H_A \right) \\ 
        &\quad +c^{BC}(\nabla_C\nabla_AH_B +\nabla_C\nabla_B H_A  - \nabla_A\nabla_B H_C) \\ 
        &\quad - \nabla_A C^{BC}\nabla_B H_C + \nabla^B C_{AB}\nabla^C H_C.
    \end{aligned}
\end{equation}
This equation reveals features hidden at first sight in equations (\ref{eq:Carrollian-evolution}),(\ref{geometriceqns}). Specifically, the non-linear structure of the underlying equation for Hajicek field (or the momentum flux) and a relevant
consequence. Considering derivatives scale as $\nabla \sim  1/L$, where $L$ is the relevant scale, as
\begin{equation}
    \dot{H} = \frac{H}{L^2}(1+c) + \frac{H}{m L}(H + m c/L).
\end{equation}
It is tempting to propose a Reynolds number defined as 
\begin{equation}
    {\rm Re} = \frac{H L/m + c}{1+c}.
\end{equation}
which bears a close resemblance to the standard expression
in fluids as well as the one suggested in~\cite{Yang:2014tla}. A
new ingredient here is the presence of $c_{AB}$ which accounts for
its role in defining the volume on which the dynamics develops.

\subsection{Monitoring quantities}\label{sec:monitor}

We will focus on a few quantities to explore the horizon's (=fluid's)
behavior. First, the black hole's area. This can be obtained from the total expansion by direct integration
(recall that from the Carrollian fluid dual perspective the expansion is the fluid's energy density). We also monitor the expansion and its angular structure. 

Out of the evolution variables \{$\theta, h_A, H_A, c_{AB}, C_{AB}$\}
we will monitor the re-summed combinations $\mathcal{H}_A = h_A+ \epsilon H_A$ and $\mathcal{C}_{AB} = c_{AB} + \epsilon C_{AB}$. Notice that while in the unperturbed regime, $\mathcal{H}_A$ is a  gauge choice, in the non-trivial regime its dynamics can help discern non-linear behavior, as argued in the previous subsection. In particular, a memory imprint will
result.
In the case of relativistic or Navier-Stokes hydrodynamics, 
the vorticity associated to the fluid's velocity is quite informative. In a Carrollian
fluid however, the dual to such velocity is the null vector $l^a$ which, by construction
does not have vorticity and by virtue of it being null, does not carry a similar physical content. 
However the Carrollian acceleration $\pi_A$ or the Hajicek field ${\cal H}^A$ provide
a sense of energy flux.
Thus, and recalling in the Eckart frame such flux governs, in particular, entropy growth; we also monitor the behavior of the vorticity associated to $\mathcal{H}_A$, defined as 
\begin{equation}
    \omega_{AB} = \partial_{[A}\mathcal{H}_{B]}.
\end{equation}
This vorticity coincides with the Carrollian vorticity defined in~\cite{Donnelly:2020xgu}. Moreover it also coincides with the Carrollian vorticity defined for a Carrollian fluid in~\cite{ciambelli_covariant_2018} up to a term that is sub-leading in perturbative order.
Further, equation (\ref{eq:viscous}) highlights the role of the Hajicek field's shear as a viscous contribution.
Geometrically, the vorticity is given by the imaginary part of the Newman-Penrose $\Psi_2$. The duality motivates an interesting physical interpretation to a purely geometrical quantity and has also been highlighted in the context of asymptotically
AdS spacetimes and the fluid-gravity correspondence on the spacetime timelike
boundary~\cite{Eling:2013sna,Green:2013zba}.

\section{Numerical Implementation}\label{sec:numerics}

We evolve equations \eqref{eq:full-eqs-first}, \eqref{eq:full-eqs-second} using the method of lines by implementing a 4th order Runge-Kutta time integrator (in Python), with time-spacing $h$. 
We use a (spectral) spin-weight decomposition to discretize the angular directions. The system is initialized at $v = v_0$ sufficiently late in time where we impose the final conditions given by $\theta^{(l)} = \mathcal{H} = 0$. In terms of spin-weighted spherical harmonics, we express the main variables as:
\begin{equation}
    \begin{aligned}
    \vartheta &= \sum_{\ell, m}\vartheta_{\ell, m}(v) Y_{\ell, m}^{(0)}(x^A), \\
    \mathcal{H} &= \sum_{\ell, m}(h_{\ell, m} + \epsilon H_{\ell,m})(v) Y_{\ell, m}^{(1)}(x^A), \\
    \mathcal{C} &= \sum_{\ell, m}(c_{\ell, m} + \epsilon C_{\ell,m})(v) Y_{\ell, m}^{(2)}(x^A).
    \end{aligned}
\end{equation}
The spatial derivatives, which we already have written in terms of spin raising and lowering operators, take a very simple form since~\cite{goldberg1967spin}:
\begin{equation}
    \begin{aligned}
        \eth Y_{\ell, m}^{(s)} = \sqrt{(\ell-s)(\ell+s+1)}Y_{\ell, m}^{(s+1)}, \\
        \bar{\eth} Y_{\ell, m}^{(s)} = -\sqrt{(\ell+s)(\ell-s+1)}Y_{\ell, m}^{(s-1)}.
    \end{aligned}
\end{equation}
Then, the system of equations becomes a system of ODEs for each angular mode of each variable. The coefficients in that system are given by integrals of three spin-weighted spherical harmonics, which can be efficiently computed from the $3j$ symbols:
\begin{equation}
\begin{aligned}
    &\int_{S^2} Y^{(s_1)}_{\ell_1 m_1}Y^{(s_2)}_{\ell_2 m_2}Y^{(s_3)}_{\ell_3 m_3} = \sqrt{\frac{(2\ell_1+1)(2\ell_2+1)(2\ell_3+1)}{4\pi}} \\
    &\times\begin{pmatrix}
    \ell_1 & \ell_2 & \ell_3 \\ 
    m_1 & m_2 & m_3 
    \end{pmatrix} \begin{pmatrix}
    \ell_1 & \ell_2 & \ell_3 \\ 
    -s_1 & -s_2 & -s_3 
    \end{pmatrix},
\end{aligned}
\end{equation}
where $s_1+s_2+s_3 = 0$. We use the implementation of the $3j$ symbols in the \textsc{sympy} package~\cite{virtanen2020scipy} to efficiently compute these integrals. The angular integrals on the sphere required to construct scalars such as the area are calculated using the \textsc{quadpy} package, and we use the \textsc{numba} package~\cite{lam2015numba} to speed--up the simulations by writing optimized machine code. The complete implementation makes use of the \textsc{numpy} framework ~\cite{harris2020array}. 

\begin{figure}
    \centering
    \includegraphics[width = 0.6\textwidth]{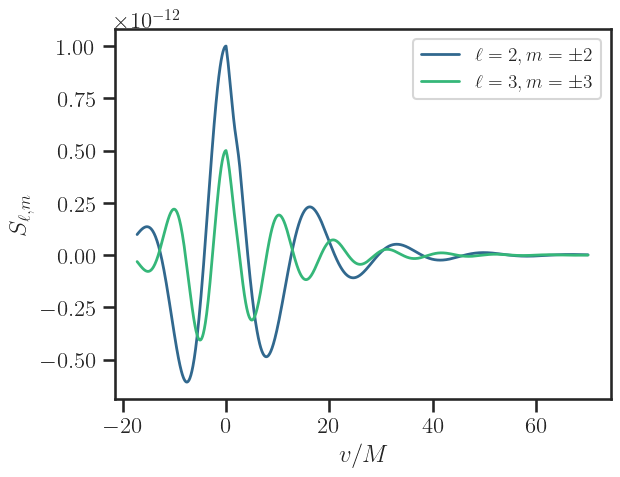}
    \caption{Source modes with excitation coefficients $s_{2,\pm 2} = 2s_{3,\pm 3}= 10^{-10}$. The source is asymmetric in time, since the initial growth is governed by a Gaussian pulse whereas the late time behavior by quasinormal modes.}
    \label{fig:Source}
\end{figure}

Finally we consider a source term describing an in-coming perturbation pulse followed by a decay motivated by quasinormal modes of a Schwarzschild black hole~\cite{Kokkotas:1999bd}. This is given, in detail, by
\begin{equation}
    s =\sum_{\ell, m} s_{\ell, m} \cos(\omega_{\ell} v) \times \begin{cases}
    \exp\left(-\frac{(v-v_S)^2}{2\sigma^2}\right), \quad &v<v_S, \\
    \exp\left(-\tau_{\ell} (v-v_S)\right) + f_{\ell}(v), \quad &v\geq v_S
    \end{cases},
\end{equation}
where we introduce an auxiliary function 
\begin{equation}
    f_{\ell}(v) = \frac{\sin\left[(v_S-v)\tau_\ell \cos(\omega_\ell v_S)\right]}{\cos(\omega_\ell v)}\exp\left(v-v_S\right)^4,
\end{equation}
in order to make the source term $C^1$. Here $v_S$ is the impinging time of the source, which for simplicity we will always set to zero $v_S = 0$, $\sigma$ the width of the Gaussian, $\omega_{\ell}$ and $\tau_{\ell}$ the real and imaginary part of the quasinormal mode frequencies, respectively, and $s_{\ell, m}$ the excitation coefficients. The explicit source profile is shown in figure~\ref{fig:Source}.
Notice the source chosen interacts with the black hole on a timescale
commensurate with its size, as a result one is not in the adiabatic regime
and modes absent in the source can be populated by non-linear interactions
(see e.g.~\cite{Sberna:2021eui}).\\
The system of equations present two practical numerical challenges. The first relates to their character which contains both first and second spatial derivatives which constrain $\Delta t$ rather strongly. The second one relates to the ``aliasing problem'', which can contaminate relevant low modes of the solution with high modes aliased down to lower modes. To address this, we adopt the standard ``two--thirds'' rule. That is, we regard the highest $1/3 \ell_{\textrm{max}}$ modes as ``superflous'' in a sense, zero-ing them at the end of each time step. 

We observe that the Eqs.~\eqref{eq:full-eqs-second} are only non-linear in the first order in perturbations variables $\{h, c\}$. For these variables, only the $(2,\pm 2)$ and the $(3, \pm 3)$ modes are excited, since those are the modes directly excited by the source. As a consequence, only modes coming from quadratic terms in these variables will be excited in our problem: the higher $\ell$ modes will appear with $\ell = 6$. We will fix then $\ell_{\textrm{max}} = 6$, and include an additional $3$ ghost modes to avoid aliasing errors. We have checked explicitly that increasing the number of ``physical'' modes does not change the solution. 

\begin{figure}[h!]
    \centering
    \includegraphics[width = 0.6\textwidth]{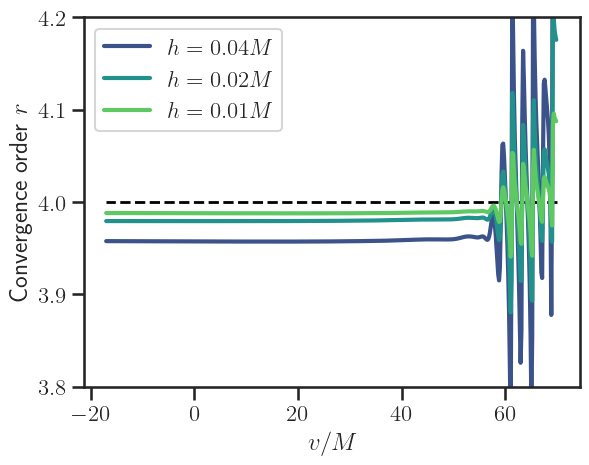}
    \caption{Convergence order measured with $h = 0.04M, 0.02M, 0.01M$ and $\ell_{\textrm{max}}=6$ with an additional $3$ ghost modes. The result is consistent with the expected 4th-order rate. }
    \label{fig:Convergence-h}
\end{figure}

We also check the convergence of the time--integrator in figure~\ref{fig:Convergence-h}. We compute the Cauchy convergence order through $r = \log_2 ( \lVert \vartheta_{2h} - \vartheta_{h}\rVert_2 / \lVert \vartheta_{h} - \vartheta_{h/2}\rVert_2)$ and the result obtained, consistent with 4th-order convergence, is shown in figure. \ref{fig:Convergence-h}
Most of the simulations presented here use an initial step--size of $h = 0.02M $, and excitation coefficients given by $s_{2,\pm 2} = \epsilon$, $s_{3, \pm 3} = 0.5 \times \epsilon$. The source is convoluted with a Gaussian profile to smoothen it out, attaining its maximum amplitude at $v_{S} = 0$ with standard deviation $\sigma =5M$.

\section{Numerical Results}\label{sec:results}
\begin{figure}
    \centering
    \includegraphics[width = 0.6\textwidth]{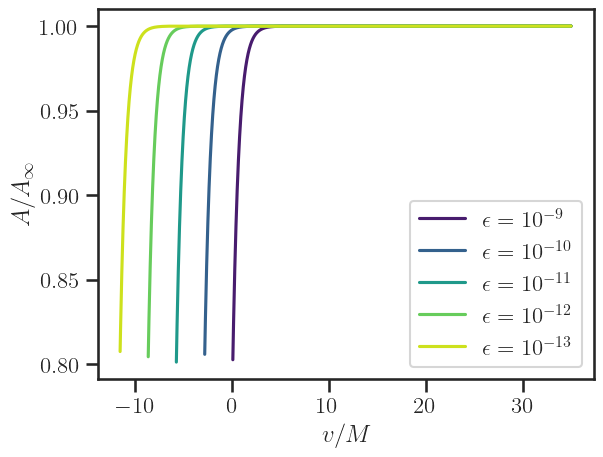}
    \caption{Evolution of the area of the black hole (normalized with respect to the area at the time where the evolution starts $v = 70M$) for different values of $\epsilon$. The perturbative scale $\epsilon$ is controlled by the amplitude of the excitation coefficients of the source $s_{\ell,m}$.}
    \label{fig:Area}
\end{figure}

The first quantity to note is the evolution of the area, which to second 
perturbative order is given by 
\begin{equation}
    \frac{dA}{dv} = 2\epsilon^2 \int d^2x \sqrt{q} \vartheta.
\end{equation}
Now we recall that the dynamics of $\vartheta$ is governed by the evolution of the first order terms $\{h, c\}$. The solution to equations~\eqref{eq:full-eqs-first} have
exponential growing modes in both directions of evolution. Since we here
demand $\theta \rightarrow 0$ in the asymptotic future, such modes are only
relevant to the past.  Their growth, in particular, will break the approximation
adopted here, so we define a ``stoppage'' time $v_0$ as the time where the area of the black hole is $80\%$ of its asymptotic (future) value. figure~\ref{fig:Area} illustrates this behavior.

\begin{figure}
    \centering
    \includegraphics[width = \textwidth]{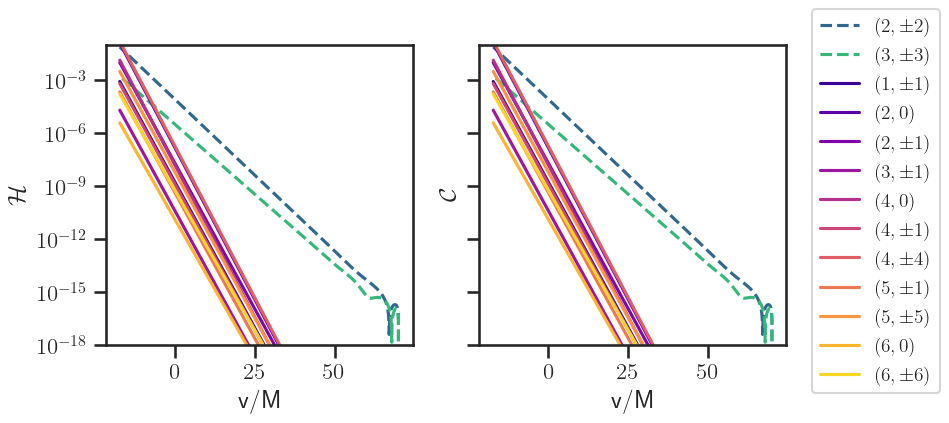}
    \caption{Modes evolution i the dynamical variables $\mathcal{H} = h + \epsilon H$ (left) and $\mathcal{C} = c + \epsilon C$ (right). The dashed lines correspond to the modes that are directly excited by the source ($(2,\pm 2)$ and $(3, \pm 3)$) whereas the solid lines correspond to modes  excited to second-order. The perturbative scale is $\epsilon = 10^{-12}$ in this case.}
    \label{fig:Variables}
\end{figure}

\begin{figure}
    \centering
    \includegraphics[width = 0.6\textwidth]{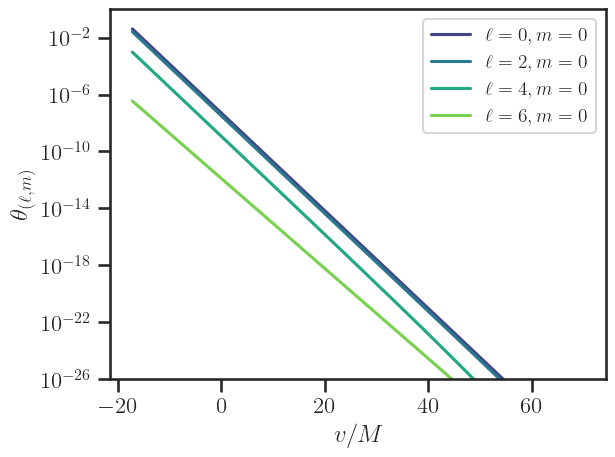}
    \caption{Modes evolution of $\theta$. The perturbative scale is $\epsilon = 10^{-12}$ in this case. The $(0,0)$ mode drives the black hole's area change. }
    \label{fig:Expansion}
\end{figure}

To better appreciate the resulting dynamics, we show the mode-by-mode contribution to the total Hajiceck field $\mathcal{H} = h + \epsilon H$ and the metric perturbation $\mathcal{C} = c + \epsilon C$ in figure~\ref{fig:Variables}, as well as the evolution of the expansion $\vartheta$ (dual to the energy density of the fluid) in figure~\ref{fig:Expansion}. We observe modes with different $(\ell,m)$ than the ones directly excited by the source are populated. 
This behavior is a consequence of the non-linear character of Eq.~\eqref{eq:full-eqs-second}. Note that close to our stopping the evolution,
non-linear modes become comparable to main (linear) perturbing ones.

\begin{figure}
    \centering
    \includegraphics[width = \textwidth]{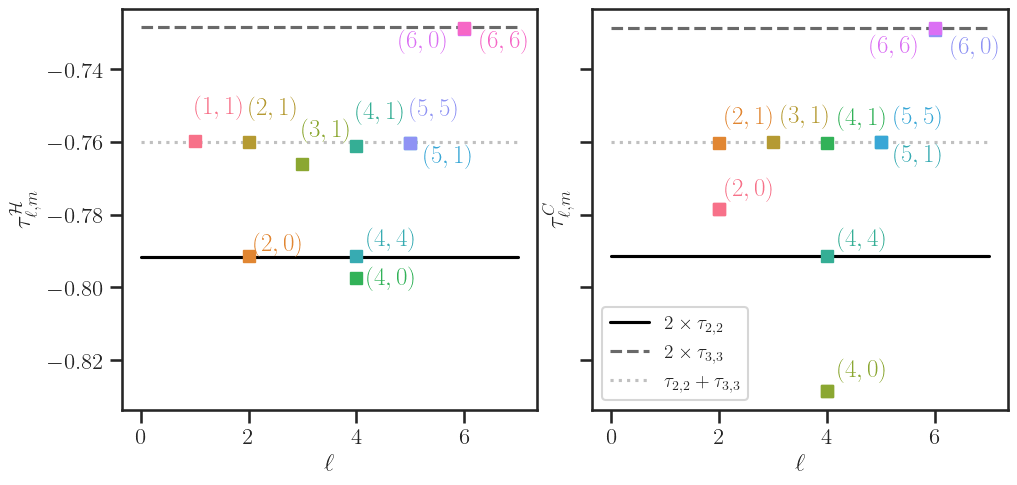}
    \caption{Decay rates for the Hajiceck degrees of freedom (left) and the metric perturbation (right) for the angular modes. The lines denote the expectation for the decay rates of the non-linearly excited modes, $2\tau_{(2,2)}$ (black, solid), $2\tau_{(3,3)}$ (dark gray, dashed), $\tau_{(2,2)} + \tau_{(3,3)}$ (lighter gray, dotted). Each square denotes the position of a mode, as indicated in the figure. }
    \label{fig:DecayRates}
\end{figure}

We measure the decay rate $\tau^{\mathcal{H}/\mathcal{C}}_{\ell,m}$ for each mode both for the Hajiceck and the angular metric perturbation degrees of freedom. One would expect a clear genealogy of the modes, where the presence of a $(2,2)$ mode in the linear perturbation $h$ to $\mathcal{H}$ induces quadratic contributions, e.g. a $(4,4)$ mode. This genealogy, in turn, manifests in the decay rates. For instance,
 $\tau_{(4,4)} = 2\tau_{(2,2)}$
(see also~\cite{Nicasio:2000ge,Zlochower:2003yh,Ripley:2020xby,Mitman:2022qdl}). Indeed, we show in figure~\ref{fig:DecayRates} that this is the case; the decay rates for extracted modes are consistent with the intuitive expectation with the exception of the
$(4,0)$ mode. 

\begin{figure}
    \centering
    \includegraphics[width = 0.6\columnwidth]{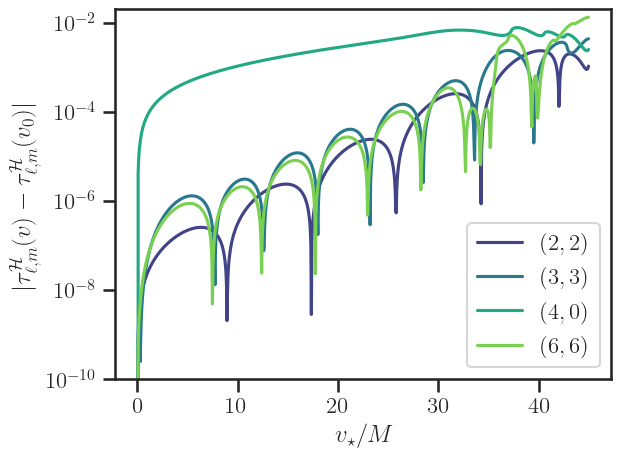}
    \caption{Dependence of the decay time $\tau_{(\ell,m)}^{\mathcal{H}}$ as a function of the final time of the fit $v_\star$, for different modes, as indicated in the legend.}
    \label{fig:TimeFit}
\end{figure}

To confirm the obtained values are largely insensitive to the time-window
used for the rates extraction, we vary it $v\in[v_S, v_\star]$, where $v_S$ is the time at which the source attains its maximum $v_\star$ the final time included in the linear regression. Figure~\ref{fig:TimeFit} illustrates the decay times measured are consistent in such window with the exception of the $m = 0$ mode. This is related to the fact that the observed $\tau^{\mathcal{H}}_{(4,0)} \neq 2\tau_{(2,2)}$. Notice also the the non-linearly excited are essentially
in lockstep with the linear ones.

\begin{figure}
    \centering
    \includegraphics[width = 0.6\columnwidth]{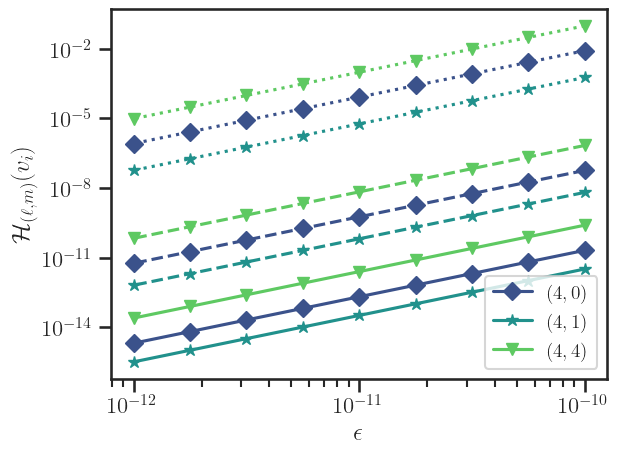}
    \caption{Modes of $\mathcal{H}$ for different $(\ell, m)$  at times $v/M = 20, 10, 0$ (solid, dashed, dotted). }
    \label{fig:ModePopulation}
\end{figure}

\begin{figure}
    \centering
    \includegraphics[width = 0.6\columnwidth]{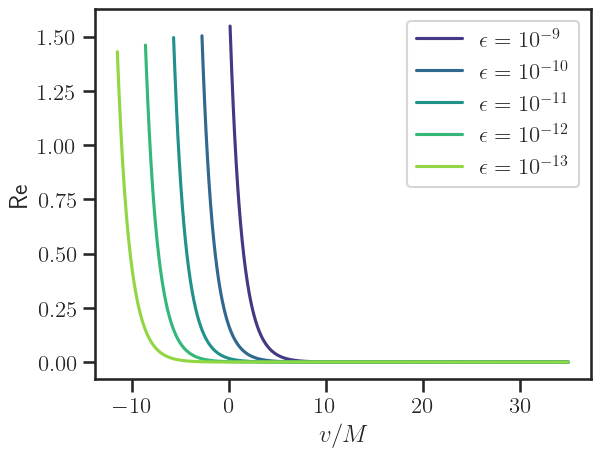}
    \caption{Evolution of the Reynolds number defined in Eq.~\eqref{eq:Reynolds-Computational} for different perturbative scales $\epsilon$. In all cases we see that the Reynolds number becomes larger than $1$ before we stop the evolution.}
    \label{fig:Reynolds}
\end{figure}

For further scrutiny, we extract the amplitude of different modes vs. time and their
dependency on the strength of the perturbation ($\epsilon$). Figure~\ref{fig:ModePopulation} illustrates some relevant modes of the Hajiceck field $\mathcal{H}$ extracted at $3$ different times.
Interestingly, at the order we work, there is no qualitatively distinct behavior in the
scaling of the non-linear modes in the ``decaying'' stage of the 
perturbation considered as the strength of the perturbation is
changed. Also, we note that higher values of $m$ have
a somewhat larger amplitude. At even earlier times however,
stronger perturbations imply a more rapid failure of the perturbative
expansion as already evident in figure~\ref{fig:Area}. We can also
connect this behavior with our proposed ``Reynolds'' number. To
evaluate it, and inspect different modes, we compute it as 
\begin{equation}\label{eq:Reynolds-Computational}
    \begin{aligned}
        \mathrm{Re} &= \sum_{\ell, m}\lvert \mathrm{Re}_{\ell,m}\rvert, \\
        \mathrm{Re}_{\ell,m} &= \frac{\mathcal{H}_{\ell,m}/\ell + \mathcal{C}_{\ell,m}}{1 + \mathcal{C}_{\ell,m}},
    \end{aligned}
\end{equation}
where we have taken as characteristic wavelength $L = 1/\ell$ for each mode, and recalled that $\mathcal{H} = h + \epsilon H$ and $\mathcal{C} = c+\epsilon C$ are the re-summed versions of the perturbations to the Hajiceck vector field and the angular metric. This Reynolds number also shows an exponential behaviour, and becomes larger than $\mathrm{Re} > 1$ prior to the stop the evolution as
shown in figure~\ref{fig:Reynolds}. This 
also indicates non-linear effects become important during the evolution. 

\begin{figure*}[t!]
    \centering
    \includegraphics[width = \textwidth]{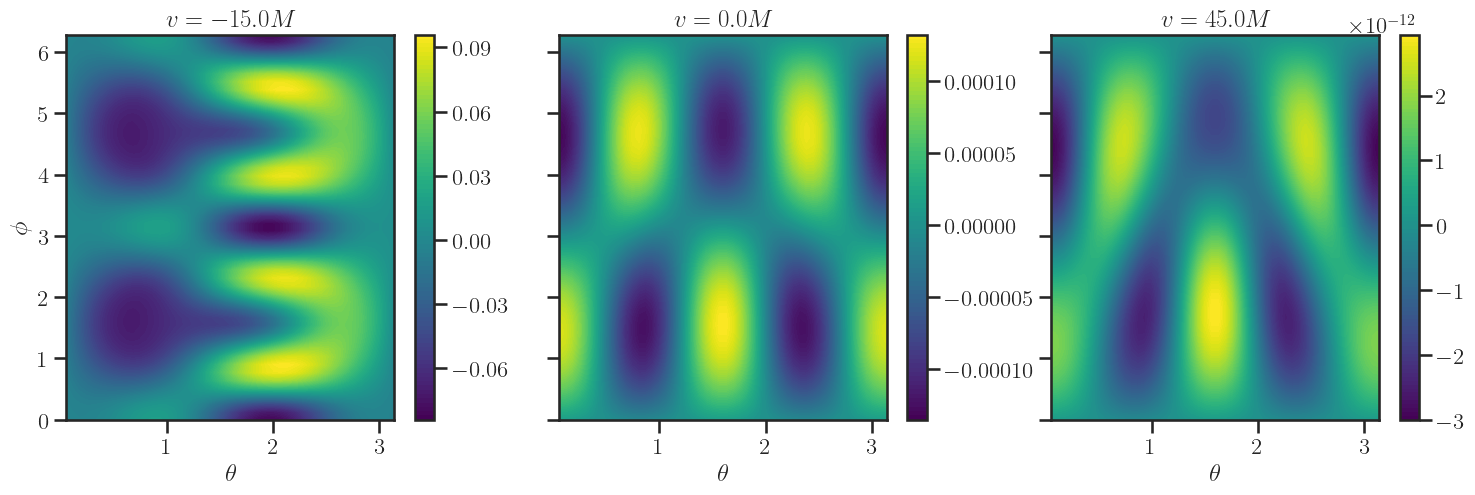}
    \caption{Evolution of the angular structure of the vorticity $2\omega = q^A\Bar{q}^B\omega_{AB}$ as a function of the angular coordinates $(\theta, \phi)$ at different times. The colormap (not normalized at different times) represents the value of the vorticity at each point.}
    \label{fig:Vorticity}
\end{figure*}

{The impact on non-linearities also reflects in changes to vorticity ($\omega_{AB}$) structure. 
Figure~\ref{fig:Vorticity} illustrates how at late times the (spin weighted) vorticity field weakens, and its angular structure is dominated by the $(2,2)$ mode. However at earlier times it becomes stronger and the presence of higher $\ell$ modes induces a more complicated angular structure. This behavior is akin to the one observed in perturbed Schwarzschild AdS black holes within studies of fluid-gravity correspondence~\cite{Green:2013zba}.}
Throughout the evolution total integral of the vorticity $\int_{S^2} \omega_{AB} = 0$ vanishes, as noted in appendix~\ref{App:Vorticiti-Enstrophy}. 

\begin{figure*}[t!]
    \centering
    \includegraphics[width = 0.95\textwidth]{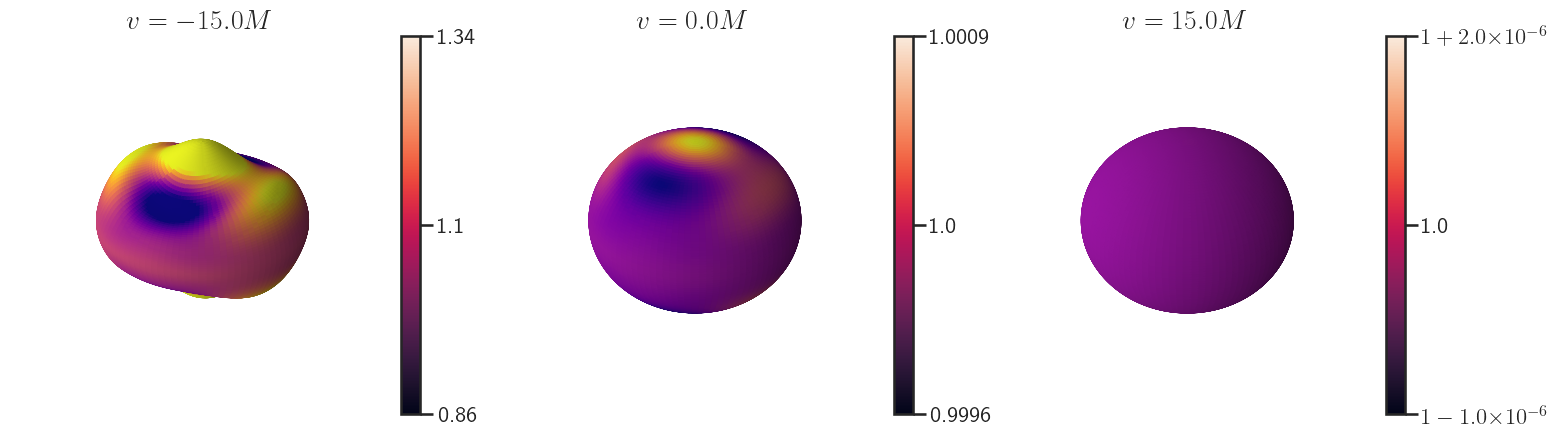}
    \caption{Isometric embedding diagram of the induced metric in the horizon at different times. The colormap represents the radius of the embedding sphere (normalized). Short wavelength perturbations become relevant at early times.}
    \label{fig:Embedding}
\end{figure*}

Finally, for a more graphical way to visualize the impact of
these non-linear features,  we show an explicit embedding diagram of the horizon in figure~\ref{fig:Embedding}. In order to do so we find an isometric embedding of the metric $q_{AB} +\epsilon \mathcal{C}_{AB}$ in $3$-dimensional space, assuming that $\epsilon \ll 1$. At late times the horizon becomes gradually more spherical, as imposed by the teleological conditions. At early times, however, it shows features with shorter wavelengths (high $\ell$ modes). As mentioned, our perturbative scheme is not able to fully capture the earlier dynamics of these modes. However, it is clear
that for the kind of perturbations here considered (perturbations with a characteristic timescale commensurate with the black hole crossing time), these short-wavelength features naturally arise. A full non-linear analysis adapted to the null surface
of the horizon, akin to the one proposed in~\cite{Gomez:2002ev} could be ideal to
further explore these observations.

As a final comment, we note the non-linear modes generated depend non-trivially
on the perturbation considered, as the number of ``channels'' for non-linear interactions depend on this.
For instance, when the perturbation is just 
a $(l,m)=(2,0)$ mode, the most relevant mode non-linearly generated is 
the $(4,0)$ though it is appreciably
smaller than, e.g. the (4,4) mode generated with an equal amplitude 
combination of a (2,2) and (2,-2)
perturbation at similar stages of the evolution. In particular
the value of $\rm Re$ is $<1$ even up to the point when we stop our
simulation in the former, while it becomes $>1$ in the latter.

\section{Final Comments}\label{sec:conclusions}
In this work we have taken first steps to explore dynamical aspects of
black holes and Carrollian fluids with a two-fold goal:\\

On one hand, 
to explore Carrollian dynamics in settings relevant to event horizons. In this
context, we identified several salient features: (i) the fluid evolves
in a time-dependent domain that is affected by the fluid's dynamics,
and higher order (in derivatives) contain both sources of dissipation
and energy, the latter connected to the behavior of the embedding space.
(ii) thermalization is linked to teleological boundary conditions. Both
a negative bulk viscosity, and a ``wrong'' sign of dissipative terms
in the evolution of the Hajicek flow appear as dooming the dynamics.
However, this is just a manifestation of the ``instability'' associated
to null surfaces, which typically will either expand to infinity or
contract towards the singularity. Certainly, the above considerations imply
that the insights drawn for Carrollian fluids in the gravitational context arguably apply
to a rather special, sub-class of fluids with such a symmetry.\footnote{ 
For other examples, see e.g. ~\cite{deBoer:2021jej} where
the equation of state, in particular, results significantly different.}
Last, and as noted in the context
of fluid-gravity correspondence, the full Einstein equations can potentially
provide guidance for a completion of Carrollian hydrodynamics, 
in the regime of an ``open system'' with energy exchange with the embedding
spacetime and backreacting on the volume in which it evolves.

On the other hand, to explore concepts of hydrodynamics and how they
can motivate new insights on the non-linear behavior of horizons.
In particular, the connection between rate of energy change, vorticity
and the flow of energy towards IR/UV regimes;
and definition of Reynold's number. In the former, we find a relation can
be established though it is unclear what message can be derived from it
as it is certainly different from the one found in, e.g. Navier-Stokes
equations and standard results can not be assumed~\cite{Kraichnan_1980}.
For the latter, we can use similar qualitative arguments as in the fluid
case to motivate a Reynolds number for perturbed black holes.
However, potential connections with phenomena understood in hydrodynamics must be 
taken with care. Negative bulk viscosity indicate instabilities on the fluid side
while in the context of black hole horizon such behavior is ruled out through teleological
boundary conditions. As well, the rate of change of the Hajicek field's energy is governed
by a significantly distinct expression than what is found in hydrodynamics. Of course
this need not be surprising; here the ``fluid'' backreacts and defines the volume
in which it exists and Carrollian symmetries are different from those in standard
and relativistic hydrodynamics. Our first steps here have been focused on exploring 
first connections for future work in this area.

From a broader point of view, we were also motivated by exploring non-linear gravitational phenomena,
and its impact/imprint on dynamical black hole horizons. Our definition of Reynolds number
seem to capture when the non-linear regime becomes relevant. In addition,
our solving for the black hole dynamics backwards in time, provides a rather natural
way to assess such a regime. In particular, our perturbation included only leading modes
and integration to the past highlighted the generation of non-linear modes and how
they become commensurate with the perturbing ones. This behavior is
particularly relevant in discussions on the onset of linear behavior in 
black hole decay to equilibrium, and the role of different mechanisms and different seeds inducing non-linear mode generation, e.g.~\cite{Yang:2014tla,Sberna:2021eui,Ripley:2020xby,Mitman:2022qdl}.

\acknowledgments

The authors would like to thank Laurent Freidel, David Garfinkle, Puttarak Jai--akson, Neev Khera, Charles Marteau, Eric Poisson, Daniele Pranzetti,
Antony Speranza, Huan Yang and Celine Zwikel for useful discussions. 
LL thanks support from NSERC through a Discovery Grant and CIFAR.
The authors acknowledge support by Perimeter Institute for Theoretical Physics. Research at Perimeter Institute is supported in part by the Government of Canada through the Department of Innovation, Science and Economic Development and by the Province of Ontario through the Ministry of Colleges and Universities. JRY also acknowledges support in the last stages of this work by the VILLUM FONDEN (grant no. 37766), by the Danish Research Foundation, and under the European Union’s H2020 ERC Advanced Grant “Black holes: gravitational engines of discovery” grant agreement no. Gravitas–101052587. 

\appendix
\section{Carrollian geometry}\label{App:Carrollian}

Here we provide some additional notes on the Carrollian geometry. There is an incipient but rapidly growing literature on this topic, so we refer the more mathematically inclined reader to \cite{Bergshoeff:2022eog}, and to \cite{ciambelli_paving_nodate, puttarak_thesis} for those more interested in the emergence of this topic in connection with physics. In this appendix we focus on Carrollian geometry in $2+1$ dimensions. 

A \emph{weak Carroll structure} is given by a $2$-dimensional Riemannian manifold, $(N, \Omega_{AB})$, and a vector field $\ell^A$, which generates the time evolution. Equivalently, we can see this structure as a fiber bundle $p: M \rightarrow N$, where $\Omega_{AB} = p(q_{ab})$ is the induced metric on $N$ and $\ell^A$ is a vertical vector field in the kernel of $q_{AB}$. A \emph{Carroll structure} is a weak Carroll structure with an Ehresmann connection, this is, a consistent way of taking derivatives within the Carroll structure itself. In the case where the Carroll structure emerges from an embedded null surface in a $4$ dimensional pseudo-Riemannian manifold, choosing a connection is equivalent to choosing a null normal $n^a$ orthogonal to the generator of the null surface.  

As per usual, it is sometimes easier to work in coordinates. The metric $q_{ab}$ in the manifold $M$ admits a Randers--Papapetrou parametrization:
\begin{equation}
    ds^2 = -c^2 (\alpha dt - b_A dx^A)^2 + \Omega_{AB}dx^A dx^B,
\end{equation}
where clearly the fiber bundle structure is recovered by taking the $c \to 0$ limit. In these coordinates the invariance under Carrollian diffeomorphisms 
\begin{equation}
    t' = t'(t, x^A), \quad x' = x'(x),
\end{equation}
is manifest. The vector field $b_A$ (which would be related to the complementary null normal $n^a$) generates the connection. Notice that a Carrollian diffeomorphism modifies the spatial derivatives like 
\begin{equation}
    \partial_A \to \frac{\partial x'_A}{\partial x_B} \left(\partial_B - \frac{\partial t'}{\partial x^B}\frac{\partial t'}{\partial t}\right).
\end{equation}
Therefore, it is not hard to see that a way to construct a \emph{Carroll covariant} derivative is by defining 
\begin{equation}
    \hat{\partial}_A = \partial_A + \frac{b_A}{\alpha}\partial_t.
\end{equation}
The Christoffel symbols also need to be modified, for similar reasons. The new Christoffel symbols are given by:
\begin{equation}
    \begin{aligned}
        \hat{\Gamma}^A_{BC} &= \frac{1}{2}\Omega^{AD}\left(\hat{\partial}_C\Omega_{BD}+\hat{\partial}_B\Omega_{DC}-\hat{\partial}_D\Omega_{BC}\right) = \Gamma^A_{BC} + c^{A}_{BC}.
    \end{aligned}
\end{equation}
Thus, covariant derivatives act on, e.g. vector fields, like 
\begin{equation}
    \hat{\nabla}_A V^B = \nabla_A V^B + \frac{b_A}{\alpha}\partial_tV^B + c^B_{AC}V^C.
\end{equation}
This is the Carrollian derivative appearing in the fluid equations~\eqref{eq:Carrollian-constraints}--\eqref{eq:Carrollian-evolution}. An interesting property is that it acquires an effective torsion when acting on scalar fields:
\begin{equation}
    [\hat{\nabla}_A, \hat{\nabla}_B]\Phi = \frac{2}{\alpha}\varpi_{AB}\Phi.
\end{equation}
We refer to $\varpi_{AB}$ as the Carrollian vorticity. If we understand the vector field $b^A$ as defining a Carrollian frame, then the vorticity associated to that frame would be naturally defined as 
\begin{equation}
    \varpi_{AB} = \partial_{[A}b_{B]} + b_{[A}\varphi_{B]},
\end{equation}
where $\varphi_A$ is the acceleration, given by 
\begin{equation}
    \varphi_A = \frac{1}{\alpha}\left(\partial_t b_A + \partial_A \alpha\right).
\end{equation}

\section{On vorticity and enstrophy}\label{App:Vorticiti-Enstrophy}

\subsection{Energy of Hajicek field}
In analogy to relevant discussion in fluids described
by the Navier-Stokes equations, let us define the following ``energy'' associated to
the vector field $\mathcal{H}_{A}$. To leading order,
\begin{equation}
    \rho = \int d^2 x \sqrt{\Omega} h_A h^A.
\end{equation}
Its rate of change is given (to leading order) by 
\begin{equation}
    \begin{aligned}
    &\dot{\rho} = 2\int d^2x \sqrt{\Omega}h^A\dot{h}_A  \\
    &= -\frac{1}{2m}\int (h^AD^2h_A + h^A D_BD_Ah^B) + \frac{1}{4m}\int h^AD_A D^BD^Cc_{BC} + \frac{1}{2m}\int h^A D^B c_{AB}  \\
    &= \frac{1}{2m}\int (D_Ah_B)^2 - \frac{1}{2m} \int (h^AD_AD_Bh^B + h^Ah_A) 
     + \frac{1}{4m}\int h^AD_A D^BD^Cc_{BC} + \frac{1}{2m}\int h^A D^B c_{AB} \\
    &= -\frac{1}{2m}\rho + \frac{1}{2m}\int \left[(D_Ah_B)^2 + (D^Ah_A)^2 \right.
     - \left.\frac{1}{2}D^Ah_A D^BD^C c_{BC} - c_{AB}D^Ah^B\right].
    \end{aligned}
\end{equation}
In the case of Navier-Stokes equations, the rate of change of the energy of the fluid flow can be related to the enstrophy, defined in terms of the flow's vorticity $w_{AB}$ as 
\begin{equation}
    Z = \int d^2 x \sqrt{\Omega} w_{AB}w^{AB}.
\end{equation}
In the context of horizon's dynamics, two proposed definitions for the vorticity
have been presented. One coincides with the imaginary part of the Newman-Penrose scalar $\Psi_2$~\cite{Donnelly:2020xgu}. The other one, proposed in~\cite{ciambelli_flat_2018} includes a further contribution so as 
to make it Carroll-covariant. However, both expressions coincide at order $\epsilon^2$. For completeness, these are given by 
\begin{equation}
\begin{aligned}
    w_{FS} &= \epsilon\partial_{[A}h_{B]} + \epsilon^2 \partial_{[A}H_{B]}, \\
    \varpi_{C} &= \epsilon\partial_{[A}h_{B]} + \epsilon^2 (\partial_{[A}H_{B]} -4m h^A \dot{h}_A).
\end{aligned}
\end{equation}
Thus,%
\begin{equation}
    Z = \left[ \int (D_A h_B)^2 - (D_A h_B)(D^B h^A)\right].
\end{equation}
Plugging this into the evolution of the energy density $\rho$ yields:
\begin{equation}
\begin{aligned}
    \dot{\rho} &= -\frac{1}{2m}\rho + \frac{1}{2m} Z + \frac{1}{2m}\int \left[\frac{}{}(D^Ah_A)^2 + (D_Ah_B)(D^Bh^A) \right.
    \\ & \quad \quad \quad \quad \quad \quad \quad \quad \quad \quad \left.- \frac{1}{2}D^Ah_A D^BD^C c_{BC} - c_{AB}D^Ah^B\right].
\end{aligned}
\end{equation}
This bound is not akin to the one found in the Navier-Stokes case. For instance, in the incompressible case, one arrives at
$\dot \rho = -2 \nu Z$ (with $\nu$ the kinematic viscosity). 
Could this be interpreted as the fluid having a negative kinematic viscosity?
An analog expression can be derived for an energy associated to
the Carrollian field $\pi_A$.
The implications of these relations are yet unclear, their consequences will be explored elsewhere.

\subsection{Vorticity conservation}

Notice that the Hajiceck equation \eqref{eq:Ray-Damour-eqs} implies that
\begin{equation}\label{eq:rotational-hajiceck}
    \partial_v(\nabla_{[A} \mathcal{H}_{B]}) = - \nabla_{[A} (\theta^{(l)}\mathcal{H}_{B]} + \nabla^C N_{B]C}).
\end{equation}
Thus, the integral of the vorticity remains constant; since we adopt $ \mathcal{H}_{A}=0$ asymptotically, it implies this integral vanishes. We have checked this condition is
satisfied in our numerical solution.

Last, $w_{AB}$ is not a Carrollian $2$-tensor, but a Carrollian vorticity can be defined, %
\begin{equation}\label{eq:Cvorticity-def}
    \varpi_{AB} = \hat{\nabla}_{[A}\mathcal{H}_{B]}.
\end{equation}
The conservation law \eqref{eq:rotational-hajiceck} is still satisfied once every derivative is promoted to its Carrollian counterpart, and this quantity has the additional advantage of being the ultrarrelativistic limit of the vorticity of a relativistic fluid \cite{ciambelli_covariant_2018}.

\bibliographystyle{jhep}
\bibliography{biblio}

\providecommand{\href}[2]{#2}\begingroup\raggedright\begin{thebibliography}{10}

\bibitem{damour_black-hole_1978}
T.~Damour, \emph{Black-hole eddy currents},
  \href{https://doi.org/10.1103/PhysRevD.18.3598}{\emph{Phys. Rev. D}
  {\bfseries 18} (1978) 3598}.

\bibitem{Thorne:1986iy}
K.S.~Thorne, R.H.~Price and D.A.~Macdonald, eds., \emph{{Black holes: the
  membrane paradigm}} (1986).

\bibitem{price_membrane_1986}
R.H.~Price and K.S.~Thorne, \emph{Membrane viewpoint on black holes:
  {Properties} and evolution of the stretched horizon},
  \href{https://doi.org/10.1103/PhysRevD.33.915}{\emph{Phys. Rev. D} {\bfseries
  33} (1986) 915}.

\bibitem{kovtun_holography_2003}
P.~Kovtun, D.T.~Son and A.O.~Starinets, \emph{Holography and hydrodynamics:
  diffusion on stretched horizons},
  \href{https://doi.org/10.1088/1126-6708/2003/10/064}{\emph{J. High Energy
  Phys.} {\bfseries 2003} (2003) 064}.

\bibitem{Baier:2007ix}
R.~Baier, P.~Romatschke, D.T.~Son, A.O.~Starinets and M.A.~Stephanov,
  \emph{{Relativistic viscous hydrodynamics, conformal invariance, and
  holography}},
  \href{https://doi.org/10.1088/1126-6708/2008/04/100}{\emph{JHEP} {\bfseries
  04} (2008) 100} [\href{https://arxiv.org/abs/0712.2451}{{\ttfamily
  0712.2451}}].

\bibitem{Bhattacharyya:2007vjd}
S.~Bhattacharyya, V.E.~Hubeny, S.~Minwalla and M.~Rangamani, \emph{{Nonlinear
  Fluid Dynamics from Gravity}},
  \href{https://doi.org/10.1088/1126-6708/2008/02/045}{\emph{JHEP} {\bfseries
  02} (2008) 045} [\href{https://arxiv.org/abs/0712.2456}{{\ttfamily
  0712.2456}}].

\bibitem{VanRaamsdonk:2008fp}
M.~Van~Raamsdonk, \emph{{Black Hole Dynamics From Atmospheric Science}},
  \href{https://doi.org/10.1088/1126-6708/2008/05/106}{\emph{JHEP} {\bfseries
  05} (2008) 106} [\href{https://arxiv.org/abs/0802.3224}{{\ttfamily
  0802.3224}}].

\bibitem{carrasco_turbulent_2012}
F.~Carrasco, L.~Lehner, R.C.~Myers, O.~Reula and A.~Singh, \emph{Turbulent
  flows for relativistic conformal fluids in 2+1 dimensions},
  \href{https://doi.org/10.1103/PhysRevD.86.126006}{\emph{Phys. Rev. D}
  {\bfseries 86} (2012) 126006}.

\bibitem{Adams:2013vsa}
A.~Adams, P.M.~Chesler and H.~Liu, \emph{{Holographic turbulence}},
  \href{https://doi.org/10.1103/PhysRevLett.112.151602}{\emph{Phys. Rev. Lett.}
  {\bfseries 112} (2014) 151602}
  [\href{https://arxiv.org/abs/1307.7267}{{\ttfamily 1307.7267}}].

\bibitem{Heller:2007qt}
M.P.~Heller and R.A.~Janik, \emph{{Viscous hydrodynamics relaxation time from
  AdS/CFT}}, \href{https://doi.org/10.1103/PhysRevD.76.025027}{\emph{Phys. Rev.
  D} {\bfseries 76} (2007) 025027}
  [\href{https://arxiv.org/abs/hep-th/0703243}{{\ttfamily hep-th/0703243}}].

\bibitem{Cremonini:2011iq}
S.~Cremonini, \emph{{The Shear Viscosity to Entropy Ratio: A Status Report}},
  \href{https://doi.org/10.1142/S0217984911027315}{\emph{Mod. Phys. Lett. B}
  {\bfseries 25} (2011) 1867}
  [\href{https://arxiv.org/abs/1108.0677}{{\ttfamily 1108.0677}}].

\bibitem{deBoer:2015ija}
J.~de~Boer, M.P.~Heller and N.~Pinzani-Fokeeva, \emph{{Effective actions for
  relativistic fluids from holography}},
  \href{https://doi.org/10.1007/JHEP08(2015)086}{\emph{JHEP} {\bfseries 08}
  (2015) 086} [\href{https://arxiv.org/abs/1504.07616}{{\ttfamily
  1504.07616}}].

\bibitem{donnay_carrollian_2019}
L.~Donnay and C.~Marteau, \emph{Carrollian physics at the black hole horizon},
  \href{https://doi.org/10.1088/1361-6382/ab2fd5}{\emph{Class. Quantum Grav.}
  {\bfseries 36} (2019) 165002}.

\bibitem{Duval:2014uva}
C.~Duval, G.W.~Gibbons and P.A.~Horvathy, \emph{{Conformal Carroll groups and
  BMS symmetry}},
  \href{https://doi.org/10.1088/0264-9381/31/9/092001}{\emph{Class. Quant.
  Grav.} {\bfseries 31} (2014) 092001}.

\bibitem{Bergshoeff:2014jla}
E.~Bergshoeff, J.~Gomis and G.~Longhi, \emph{{Dynamics of Carroll Particles}},
  \href{https://doi.org/10.1088/0264-9381/31/20/205009}{\emph{Class. Quant.
  Grav.} {\bfseries 31} (2014) 205009}
  [\href{https://arxiv.org/abs/1405.2264}{{\ttfamily 1405.2264}}].

\bibitem{Bergshoeff:2022eog}
E.~Bergshoeff, J.~Figueroa-O'Farrill and J.~Gomis, \emph{{A non-lorentzian
  primer}},  \href{https://arxiv.org/abs/2206.12177}{{\ttfamily 2206.12177}}.

\bibitem{deBoer:2017ing}
J.~de~Boer, J.~Hartong, N.A.~Obers, W.~Sybesma and S.~Vandoren, \emph{{Perfect
  Fluids}}, \href{https://doi.org/10.21468/SciPostPhys.5.1.003}{\emph{SciPost
  Phys.} {\bfseries 5} (2018) 003}
  [\href{https://arxiv.org/abs/1710.04708}{{\ttfamily 1710.04708}}].

\bibitem{Petkou:2022bmz}
A.C.~Petkou, P.M.~Petropoulos, D.R.~Betancour and K.~Siampos,
  \emph{{Relativistic fluids, hydrodynamic frames and their Galilean versus
  Carrollian avatars}},
  \href{https://doi.org/10.1007/JHEP09(2022)162}{\emph{JHEP} {\bfseries 09}
  (2022) 162} [\href{https://arxiv.org/abs/2205.09142}{{\ttfamily
  2205.09142}}].

\bibitem{Gomez:2000kn}
R.~Gomez, S.~Husa and J.~Winicour, \emph{{Complete null data for a black hole
  collision}}, \href{https://doi.org/10.1103/PhysRevD.64.024010}{\emph{Phys.
  Rev. D} {\bfseries 64} (2001) 024010}
  [\href{https://arxiv.org/abs/gr-qc/0009092}{{\ttfamily gr-qc/0009092}}].

\bibitem{chrusciel2020geometry}
P.T.~Chru{\'s}ciel, \emph{Geometry of Black Holes}, vol.~169, Oxford University
  Press (2020).

\bibitem{adami_null_2021}
H.~Adami, D.~Grumiller, M.M.~Sheikh-Jabbari, V.~Taghiloo, H.~Yavartanoo and
  C.~Zwikel, \emph{Null boundary phase space: slicings, news and memory},
  \href{https://doi.org/10.1007/JHEP11(2021)155}{\emph{J. High Energ. Phys.}
  {\bfseries 2021} (2021) 155}.

\bibitem{Wald:1984rg}
R.M.~Wald, \emph{{General Relativity}}, Chicago Univ. Pr., Chicago, USA (1984),
  \href{https://doi.org/10.7208/chicago/9780226870373.001.0001}{10.7208/chicago/9780226870373.001.0001}.

\bibitem{donnay_extended_2016}
L.~Donnay, G.~Giribet, H.A.~González and M.~Pino, \emph{Extended symmetries at
  the black hole horizon},
  \href{https://doi.org/10.1007/JHEP09(2016)100}{\emph{J. High Energ. Phys.}
  {\bfseries 2016} (2016) 100}.

\bibitem{Hartong:2015xda}
J.~Hartong, \emph{{Gauging the Carroll Algebra and Ultra-Relativistic
  Gravity}}, \href{https://doi.org/10.1007/JHEP08(2015)069}{\emph{JHEP}
  {\bfseries 08} (2015) 069}
  [\href{https://arxiv.org/abs/1505.05011}{{\ttfamily 1505.05011}}].

\bibitem{Ciambelli:2019lap}
L.~Ciambelli, R.G.~Leigh, C.~Marteau and P.M.~Petropoulos, \emph{{Carroll
  Structures, Null Geometry and Conformal Isometries}},
  \href{https://doi.org/10.1103/PhysRevD.100.046010}{\emph{Phys. Rev. D}
  {\bfseries 100} (2019) 046010}
  [\href{https://arxiv.org/abs/1905.02221}{{\ttfamily 1905.02221}}].

\bibitem{Freidel:2022bai}
L.~Freidel and P.~Jai-akson, \emph{{Carrollian hydrodynamics from symmetries}},
   \href{https://arxiv.org/abs/2209.03328}{{\ttfamily 2209.03328}}.

\bibitem{ciambelli_paving_nodate}
L.~Ciambelli, \emph{Paving the fluid road to flat holography}, {PhD}
  dissertation, Université Paris-Saclay.

\bibitem{lehner_gravitational_2016}
L.~Lehner, R.C.~Myers, E.~Poisson and R.D.~Sorkin, \emph{Gravitational action
  with null boundaries},
  \href{https://doi.org/10.1103/PhysRevD.94.084046}{\emph{Phys. Rev. D}
  {\bfseries 94} (2016) 084046}.

\bibitem{Chandrasekaran:2021hxc}
V.~Chandrasekaran, E.E.~Flanagan, I.~Shehzad and A.J.~Speranza,
  \emph{{Brown-York charges at null boundaries}},
  \href{https://doi.org/10.1007/JHEP01(2022)029}{\emph{JHEP} {\bfseries 01}
  (2022) 029} [\href{https://arxiv.org/abs/2109.11567}{{\ttfamily
  2109.11567}}].

\bibitem{York:1986it}
J.W.~York, Jr., \emph{{Black hole thermodynamics and the Euclidean Einstein
  action}}, \href{https://doi.org/10.1103/PhysRevD.33.2092}{\emph{Phys. Rev. D}
  {\bfseries 33} (1986) 2092}.

\bibitem{Brown:1992br}
J.D.~Brown and J.W.~York, Jr., \emph{{Quasilocal energy and conserved charges
  derived from the gravitational action}},
  \href{https://doi.org/10.1103/PhysRevD.47.1407}{\emph{Phys. Rev. D}
  {\bfseries 47} (1993) 1407}.

\bibitem{Freidel:2022vjq}
L.~Freidel and P.~Jai-akson, \emph{{Carrollian hydrodynamics and symplectic
  structure on stretched horizons}},
  \href{https://arxiv.org/abs/2211.06415}{{\ttfamily 2211.06415}}.

\bibitem{Gourgoulhon:2005ng}
E.~Gourgoulhon and J.L.~Jaramillo, \emph{{A 3+1 perspective on null
  hypersurfaces and isolated horizons}},
  \href{https://doi.org/10.1016/j.physrep.2005.10.005}{\emph{Phys. Rept.}
  {\bfseries 423} (2006) 159}.

\bibitem{ciambelli_covariant_2018}
L.~Ciambelli, C.~Marteau, A.C.~Petkou, P.M.~Petropoulos and K.~Siampos,
  \emph{Covariant {Galilean} versus {Carrollian} hydrodynamics from
  relativistic fluids},
  \href{https://doi.org/10.1088/1361-6382/aacf1a}{\emph{Class. Quantum Grav.}
  {\bfseries 35} (2018) 165001}.

\bibitem{Gomez:2002ev}
R.~Gomez, S.~Husa, L.~Lehner and J.~Winicour, \emph{{Gravitational waves from a
  fissioning white hole}},
  \href{https://doi.org/10.1103/PhysRevD.66.064019}{\emph{Phys. Rev. D}
  {\bfseries 66} (2002) 064019}
  [\href{https://arxiv.org/abs/gr-qc/0205038}{{\ttfamily gr-qc/0205038}}].

\bibitem{ashtekar_dynamical_2013}
A.~Ashtekar, M.~Campiglia and S.~Shah, \emph{Dynamical {Black} {Holes}:
  {Approach} to the {Final} {State}},
  \href{https://doi.org/10.1103/PhysRevD.88.064045}{\emph{Phys. Rev. D}
  {\bfseries 88} (2013) 064045}.

\bibitem{Ashtekar:2021kqj}
A.~Ashtekar, N.~Khera, M.~Kolanowski and J.~Lewandowski, \emph{{Charges and
  fluxes on (perturbed) non-expanding horizons}},
  \href{https://doi.org/10.1007/JHEP02(2022)066}{\emph{JHEP} {\bfseries 02}
  (2022) 066} [\href{https://arxiv.org/abs/2112.05608}{{\ttfamily
  2112.05608}}].

\bibitem{Loutrel:2020wbw}
N.~Loutrel, J.L.~Ripley, E.~Giorgi and F.~Pretorius, \emph{{Second Order
  Perturbations of Kerr Black Holes: Reconstruction of the Metric}},
  \href{https://doi.org/10.1103/PhysRevD.103.104017}{\emph{Phys. Rev. D}
  {\bfseries 103} (2021) 104017}
  [\href{https://arxiv.org/abs/2008.11770}{{\ttfamily 2008.11770}}].

\bibitem{Ripley:2020xby}
J.L.~Ripley, N.~Loutrel, E.~Giorgi and F.~Pretorius, \emph{{Numerical
  computation of second order vacuum perturbations of Kerr black holes}},
  \href{https://doi.org/10.1103/PhysRevD.103.104018}{\emph{Phys. Rev. D}
  {\bfseries 103} (2021) 104018}
  [\href{https://arxiv.org/abs/2010.00162}{{\ttfamily 2010.00162}}].

\bibitem{Gray:2022svz}
F.~Gray, D.~Kubiznak, T.R.~Perche and J.~Redondo-Yuste, \emph{{Carrollian
  Motion in Magnetized Black Hole Horizons}},
  \href{https://arxiv.org/abs/2211.13695}{{\ttfamily 2211.13695}}.

\bibitem{newman_approach_1962}
E.~Newman and R.~Penrose, \emph{An {Approach} to {Gravitational} {Radiation} by
  a {Method} of {Spin} {Coefficients}},
  \href{https://doi.org/10.1063/1.1724257}{\emph{Journal of Mathematical
  Physics} {\bfseries 3} (1962) 566}.

\bibitem{gomez_eth_1997}
R.~Gómez, L.~Lehner, P.~Papadopoulos and J.~Winicour, \emph{The eth formalism
  in numerical relativity},
  \href{https://doi.org/10.1088/0264-9381/14/4/013}{\emph{Class. Quantum Grav.}
  {\bfseries 14} (1997) 977}.

\bibitem{Bishop:1996gt}
N.T.~Bishop, R.~Gomez, L.~Lehner and J.~Winicour, \emph{{Cauchy-characteristic
  extraction in numerical relativity}},
  \href{https://doi.org/10.1103/PhysRevD.54.6153}{\emph{Phys. Rev. D}
  {\bfseries 54} (1996) 6153}
  [\href{https://arxiv.org/abs/gr-qc/9705033}{{\ttfamily gr-qc/9705033}}].

\bibitem{Kokkotas:1999bd}
K.D.~Kokkotas and B.G.~Schmidt, \emph{{Quasinormal modes of stars and black
  holes}}, \href{https://doi.org/10.12942/lrr-1999-2}{\emph{Living Rev. Rel.}
  {\bfseries 2} (1999) 2}.

\bibitem{Yang:2014tla}
H.~Yang, A.~Zimmerman and L.~Lehner, \emph{{Turbulent Black Holes}},
  \href{https://doi.org/10.1103/PhysRevLett.114.081101}{\emph{Phys. Rev. Lett.}
  {\bfseries 114} (2015) 081101}.

\bibitem{Donnelly:2020xgu}
W.~Donnelly, L.~Freidel, S.F.~Moosavian and A.J.~Speranza, \emph{{Gravitational
  edge modes, coadjoint orbits, and hydrodynamics}},
  \href{https://doi.org/10.1007/JHEP09(2021)008}{\emph{JHEP} {\bfseries 09}
  (2021) 008} [\href{https://arxiv.org/abs/2012.10367}{{\ttfamily
  2012.10367}}].

\bibitem{Eling:2013sna}
C.~Eling and Y.~Oz, \emph{{Holographic Vorticity in the Fluid/Gravity
  Correspondence}}, \href{https://doi.org/10.1007/JHEP11(2013)079}{\emph{JHEP}
  {\bfseries 11} (2013) 079} [\href{https://arxiv.org/abs/1308.1651}{{\ttfamily
  1308.1651}}].

\bibitem{Green:2013zba}
S.R.~Green, F.~Carrasco and L.~Lehner, \emph{{Holographic Path to the Turbulent
  Side of Gravity}},
  \href{https://doi.org/10.1103/PhysRevX.4.011001}{\emph{Phys. Rev. X}
  {\bfseries 4} (2014) 011001}
  [\href{https://arxiv.org/abs/1309.7940}{{\ttfamily 1309.7940}}].

\bibitem{goldberg1967spin}
J.N.~Goldberg, A.J.~MacFarlane, E.T.~Newman, F.~Rohrlich and E.G.~Sudarshan,
  \emph{Spin-s spherical harmonics and {\dh}}, {\emph{Journal of Mathematical
  Physics} {\bfseries 8} (1967) 2155}.

\bibitem{virtanen2020scipy}
P.~Virtanen, R.~Gommers, T.E.~Oliphant, M.~Haberland, T.~Reddy, D.~Cournapeau
  et~al., \emph{Scipy 1.0: fundamental algorithms for scientific computing in
  python}, {\emph{Nature methods} {\bfseries 17} (2020) 261}.

\bibitem{lam2015numba}
S.K.~Lam, A.~Pitrou and S.~Seibert, \emph{Numba: A llvm-based python jit
  compiler},  in \emph{Proceedings of the Second Workshop on the LLVM Compiler
  Infrastructure in HPC}, pp.~1--6, 2015.

\bibitem{harris2020array}
C.R.~Harris, K.J.~Millman, S.J.~Van Der~Walt, R.~Gommers, P.~Virtanen,
  D.~Cournapeau et~al., \emph{Array programming with numpy}, {\emph{Nature}
  {\bfseries 585} (2020) 357}.

\bibitem{Sberna:2021eui}
L.~Sberna, P.~Bosch, W.E.~East, S.R.~Green and L.~Lehner, \emph{{Nonlinear
  effects in the black hole ringdown: Absorption-induced mode excitation}},
  \href{https://doi.org/10.1103/PhysRevD.105.064046}{\emph{Phys. Rev. D}
  {\bfseries 105} (2022) 064046}.

\bibitem{Nicasio:2000ge}
C.O.~Nicasio, R.~Gleiser and J.~Pullin, \emph{{Second order perturbations of a
  Schwarzschild black hole: Inclusion of odd parity perturbations}},
  \href{https://doi.org/10.1023/A:1001994318436}{\emph{Gen. Rel. Grav.}
  {\bfseries 32} (2000) 2021}
  [\href{https://arxiv.org/abs/gr-qc/0001021}{{\ttfamily gr-qc/0001021}}].

\bibitem{Zlochower:2003yh}
Y.~Zlochower, R.~Gomez, S.~Husa, L.~Lehner and J.~Winicour, \emph{{Mode
  coupling in the nonlinear response of black holes}},
  \href{https://doi.org/10.1103/PhysRevD.68.084014}{\emph{Phys. Rev. D}
  {\bfseries 68} (2003) 084014}
  [\href{https://arxiv.org/abs/gr-qc/0306098}{{\ttfamily gr-qc/0306098}}].

\bibitem{Mitman:2022qdl}
K.~Mitman et~al., \emph{Nonlinearities in black hole ringdowns},
  \href{https://arxiv.org/abs/2208.07380}{{\ttfamily 2208.07380}}.

\bibitem{deBoer:2021jej}
J.~de~Boer, J.~Hartong, N.A.~Obers, W.~Sybesma and S.~Vandoren, \emph{{Carroll
  Symmetry, Dark Energy and Inflation}},
  \href{https://doi.org/10.3389/fphy.2022.810405}{\emph{Front. in Phys.}
  {\bfseries 10} (2022) 810405}
  [\href{https://arxiv.org/abs/2110.02319}{{\ttfamily 2110.02319}}].

\bibitem{Kraichnan_1980}
R.H.~Kraichnan and D.~Montgomery, \emph{Two-dimensional turbulence},
  \href{https://doi.org/10.1088/0034-4885/43/5/001}{\emph{Reports on Progress
  in Physics} {\bfseries 43} (1980) 547}.

\bibitem{puttarak_thesis}
P.~Jai-akson, \emph{Edge Modes and Carrollian Hydrodynamics on Stretched
  Horizons}, {PhD} dissertation, University of Waterloo.

\bibitem{ciambelli_flat_2018}
L.~Ciambelli, C.~Marteau, A.C.~Petkou, P.M.~Petropoulos and K.~Siampos,
  \emph{Flat holography and {Carrollian} fluids},
  \href{https://doi.org/10.1007/JHEP07(2018)165}{\emph{J. High Energ. Phys.}
  {\bfseries 2018} (2018) 165}.

\end{thebibliography}\endgroup
\end{document}